\documentclass[11pt,oneside,a4paper]{article}

\usepackage{xspace}%
\usepackage{tikz}
\usepackage{tikz-3dplot}
\usepackage{cite}
\usetikzlibrary{arrows.meta}
\usepackage{physics}
\usepackage{booktabs}
\usepackage[colorlinks=true,allcolors=blue]{hyperref}

\title{A spacetime model for the perturbative swing}
\author{Torbj{\"o}rn Lundberg, Leif L{\"o}nnblad}

\newenvironment{itemise}{\begin{itemize}}{\end{itemize}}
\renewcommand{\sec}[1]{
  Sec.~\ref{sec:#1}}
\newcommand{\app}[1]{
  Appendix~\ref{sec:#1}}
\newcommand{\eq}[1]{
  Eq.~\eqref{eq:#1}}
\newcommand{\fig}[1]{
  Fig.~\ref{fig:#1}}

\newcommand{\swing}{
  swing\xspace}
\def\AA{\ensuremath{AA}}
\def\pA{\ensuremath{\text{p}A}}
\def\pp{\ensuremath{\text{pp}}}
\def\pom{{\ensuremath{\text{I\!P}}}}
\def\ee{\ensuremath{\text{e}^+\text{e}^-}}
\def\swstr{\ensuremath{\alpha_{\text{sw}}}}
\def\pythia{P\textsc{ythia}\xspace}
\def\pytppp{P\textsc{ythia8}\xspace}
\def\herwig{H\textsc{erwig}\xspace}
\def\ariadne{A\textsc{riadne}\xspace}
\def\eg{e.g.\xspace}
\def\ie{i.e.\xspace}
\def\cf{c.f.\xspace}

\definecolor{brightpink}{rgb}{1.0, 0.0, 0.5}

\definecolor{darkgreen}{rgb}{0.0, 0.5, 0.0}

\begin{document}
\maketitle

\begin{abstract}
  We present the development of a new colour reconnection model in \pythia where the full spacetime separation between partons constrains the reconnection probability throughout the final-state radiative shower.
  Building upon previous ideas of a perturbative dipole-based model, our objective is to introduce realistic spatially-based colour reconnections within the Angantyr model for heavy-ion collisions.
  Previous models were limited to the use of a rudimentary separation measure in the transverse plane to control reconnections.
  We investigate the impact of this novel scheme on final states in \ee, \pp, and pPb collisions as well as in PbPb collision systems and our main goal in this paper is to achieve a reasonable description of basic multiplicities in systems of varying sizes.
\end{abstract}

{\bf Keywords:} particle physics, phenomenology, quantum chromodynamics, electron-\hspace{0pt}positron collisions, proton-proton collisions, proton-ion collisions, heavy-ion collisions, event generators, \pytppp, Angantyr, tuning, colour reconnections, perturbative swing

\section{Introduction}
\label{sec:intro}
  %
When modelling hadronic final states in high-energy hadron collisions
using multi-parton scattering models, the introduction of a
\textit{colour reconnection} mechanism has proven to be essential
in order to reproduce experimental measurements. In particular,
observables such as the average transverse momentum as a function of
charged multiplicity is virtually impossible to describe if no colour
crosstalk between different partonic subscatterings is allowed. This
was recognised already when the first multi-parton interaction (MPI) model
was introduced in the \pythia event
generator\footnote{%
  For the recent release, see \cite{Bierlich:2022pfr}.
}
in the eighties~\cite{Sjostrand:1987su}.

Colour reconnections may also occur in other
collision systems; perhaps the most clean example being electron-positron
annihilation into hadrons,
\mbox{\ee$\to\,$(W$^+\to q_1\bar{q}_2$)(W$^-\to q_3\bar{q}_4$)},
where a quark from one W decay can form a colour singlet with an
antiquark from the other.
The effect of such reconnections is expected to be small
but is nevertheless large enough to be one of the
main contributors to the theoretical systematic uncertainty in the W mass
measurement at LEP2 (see, \eg,~\cite{DELPHI:2008avl}).

Several reconnection models have been implemented in event generators
and, besides the original one in \cite{Sjostrand:1987su}, there are
a few options in \pythia. The most relevant model for the work presented
here is referred to as the \emph{QCD-based colour reconnection},
\cite{Christiansen:2015yqa}, in which \mbox{SU(3)} colour algebra is taken into
account when deciding which partons are allowed to reconnect in colour
space. This model is special in that it allows reconnections
that form non-trivial so-called \emph{colour junction} structures that carry
baryon number.
We note that the \herwig event generator, \cite{Bellm:2015jjp}, also includes a similar model
that allows for the creation of baryonic clusters~\cite{Bellm:2020fny}.

Furthermore, basic colour algebra was used in the so-called
\emph{perturbative swing model}, \cite{Lonnblad:1995yk}, implemented in the
\ariadne generator, \cite{Lonnblad:1992tz}, and we adopt the same starting point
for the work presented here.
\ariadne was the first implementation of a
dipole-based parton shower, and the perturbative swing model allowed
for the reconnection, or \emph{swing}, between the partons of pairs of
dipoles in the same colour state, thereby affecting the perturbative
stage of the event generation process.

Taking inspiration from \ariadne in this paper,
we endeavour to develop a full spacetime picture of colour
reconnections in hadronic collisions.
Our work here is conducted within the context
of recent ideas by our group in Lund, focusing on the development of a
microscopic picture of dense hadron and heavy-ion collisions systems, based on
string degrees of freedom, as an alternative to the conventional
thermodynamic modelling of a Quark--Gluon Plasma (QGP).
So far, such earlier
work has mainly dealt with non-perturbative interactions between overlapping
strings, resulting in the \textit{Shoving} model,
\cite{Bierlich:2017vhg,Bierlich:2020naj}, describing flow effects
caused by the repulsion between strings, and the \textit{Rope} model,
\cite{Bierlich:2014xba,Bierlich:2022ned}, describing how the increase
in string tension affects, \eg, the strangeness production in the
string fragmentation.
Both these models rely on a detailed spacetime
description of the colour fields and are therefore very sensitive to
how colour reconnections are modelled.

Just like most other colour reconnection models, the main result of
the swing mechanism is to lower the overall energy content of the collision system
by reducing the amount of colour field it contains.
In \pythia, this is achieved by permitting only reconnections that reduce
the overall \emph{string length}, \cite{Andersson:1988ee}, while in \herwig, the
requirement for reconnections is that cluster masses are reduced.
In \ariadne, the
reduction of dipole invariant masses is favoured;
however, unlike other models, swings that increase dipole masses are also
allowed, resulting in a kind of heat-bath algorithm that seeks to
globally minimise the invariant masses.

In all colour reconnection models, there is a significant effect on the
final-state multiplicity in hadronic collisions, and it is natural to
assume that for collisions between heavy ions, the effect on the
multiplicity would be even more pronounced. There are many more coloured partons
produced in such collisions and, thus, many more colour dipoles may
reconnect. The partons in heavy-ion collisions are, however,
spread out in spacetime in the sense that the collision region is large in
comparison to a single nucleon,
and it is not reasonable to accept that partons separated by several
femtometres should be allowed to reconnect.
Most models mentioned above use purely momentum based
pictures and are therefore not well suited to be applied in heavy-ion
collisions.

In a recent publication, \cite{Lonnblad:2023stc}, a first attempt was made
to adapt and apply the QCD-based colour reconnection model to heavy-ion
collision events generated by the Angantyr model
in~\pytppp\cite{Bierlich:2018xfw}.
In that work, a simple cut in
impact parameter was used to suppress reconnections between dipoles
that are too far apart in impact parameter space and the results demonstrated
that it should be possible to apply colour reconnections also in
heavy-ion collisions.

In this article, we will attempt to go further and develop a more
complete spacetime picture of the colour structure of events. Our
starting point will be the perturbative swing model, as we mentioned above,
which we assume
provides an accurate description at very small distances using only a
momentum-based framework. At larger distances, we will then dynamically
include the spatial separation in a way that treats all partons on equal
footing irrespective of whether they are in the soft underlying event
or in a hard jet.

This paper is organised as follows. In \sec{colo-reconn-pert}, we discuss colour
reconnections in general with an emphasis on the perturbative swing
model. Then, in \sec{spacetime-picture}, we describe how we introduce the
spacetime separation of partons in the swing model. We investigate the
effects of the implementation of the new algorithm in \sec{res} for
\ee\ annihilation into hadrons at LEP energies, and for hadronic
collisions at the LHC, encompassing \pp, pPb, and PbPb collisions. In
\sec{disc}, we finally summarise and discuss our results, as well as outline the
further developments we envisage.

\section{Colour reconnections and perturbative swing}
\label{sec:colo-reconn-pert}

Most colour reconnection models implement some procedure for
minimising the free energy of the collision system that can be used to produce
hadrons in an event.
In the case of string fragmentation, this amounts to minimising the
string length, while for cluster hadronisation, it typically means
minimising the cluster masses.\footnote{%
  In the current version of \herwig, there is an option to instead minimise a
  combination of rapidity separation and impact parameter separation of
  partons~\cite{Bellm:2020fny}.
}
The string length is typically defined in
terms of the \emph{lambda measure} which for a dipole between a $q$ and a
$\bar{q}$ in its simplest form is given by
\begin{equation}
  \label{eq:lambda}
  \lambda_{q\bar{q}} \propto \log\left(\frac{m^2_{q\bar{q}}}{m^2_0}\right),
\end{equation}
where $m_{q\bar{q}}$ is the invariant mass of the \mbox{$q\bar{q}$}-pair
and $m_0$ is a hadronic mass scale~\cite{Andersson:1988ee}.
Minimising this measure is
therefore analogous to minimising dipole or cluster masses.

We will here concentrate on the perturbative swing model
but the reconnection procedure is similar to what is
used also in other reconnection models. The main difference is that,
while other models only act on parton configurations after the parton
shower phase of the event generation, the perturbative swing is
applied continually during the final-state radiative (FSR) shower.

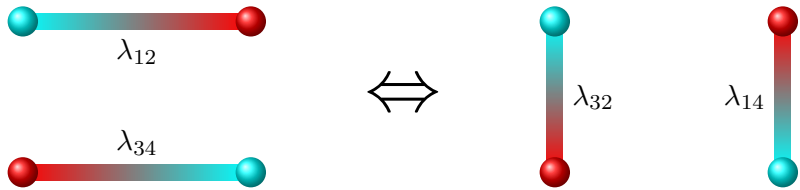
\begin{figure}[t]
  \centering
  \begin{tikzpicture}
  \coordinate (O) at (-0.5,-2.5);
  %
  \node at (0,0) {\resizebox{1cm}{!}{$\Leftrightarrow$}};
  \pgfmathsetmacro{\pOnex}{-2}
  \pgfmathsetmacro{\pOnez}{1}
  \pgfmathsetmacro{\pTwox}{-5}
  \pgfmathsetmacro{\pTwoz}{1}
  \pgfmathsetmacro{\pThreex}{-2}
  \pgfmathsetmacro{\pThreez}{-1}
  \pgfmathsetmacro{\pFourx}{-5}
  \pgfmathsetmacro{\pFourz}{-1}
  \coordinate (p1) at (\pOnex,\pOnez);
  \coordinate (p2) at (\pTwox,\pTwoz);
  \coordinate (p3) at (\pThreex,\pThreez);
  \coordinate (p4) at (\pFourx,\pFourz);
  \draw[line width=6pt,red] (p1) -- (p2) node[midway,below] {\color{black}$\lambda_{12}$};
  \draw[line width=6pt,red] (p3) -- (p4) node[midway,above] {\color{black}$\lambda_{34}$};
  \def\connect#1#2#3{%
  \path let
    \p1 = ($(#2)-(#1)$),
    \n1 = {veclen(\p1)},
    \n2 = {atan2(\x1,\y1)} 
  in
    (#1) -- (#2) node[#3, midway, sloped, shading angle=\n2+90, minimum width=\n1, inner sep=0pt, #3] {};
  }
  \def\antired{cyan}
  \connect{p1}{p2}{right color=red, left color=\antired, minimum height=6pt}
  \connect{p3}{p4}{left color=red, right color=\antired, minimum height=6pt}
  \pgfmathsetmacro{\th}{55}  
  \pgfmathsetmacro{\eOne}{3.}  
  \pgfmathsetmacro{\eTwo}{3.}  
  \pgfmathsetmacro{\eThree}{3.}  
  \pgfmathsetmacro{\eFour}{3.}  
  \pgfmathsetmacro{\radius}{.2}  
  \pgfmathsetmacro{\radNudge}{1.4}  
  \pgfmathsetmacro{\vx}{{sin(\th)}}
  \pgfmathsetmacro{\vz}{{cos(\th)}}
  \pgfmathsetmacro{\offset}{\radNudge*\radius}
  \shade[ball color = red!100, opacity = 1.] (p1) circle (\radius);
  \shade[ball color = \antired!100, opacity = 1.] (p2) circle (\radius);
  \shade[ball color = \antired!100, opacity = 1.] (p3) circle (\radius);
  \shade[ball color = red!100, opacity = 1.] (p4) circle (\radius);
  %
  %
  \pgfmathsetmacro{\pOnex}{5}
  \pgfmathsetmacro{\pOnez}{1}
  \pgfmathsetmacro{\pTwox}{2}
  \pgfmathsetmacro{\pTwoz}{1}
  \pgfmathsetmacro{\pThreex}{5}
  \pgfmathsetmacro{\pThreez}{-1}
  \pgfmathsetmacro{\pFourx}{2}
  \pgfmathsetmacro{\pFourz}{-1}
  \coordinate (p1) at (\pOnex,\pOnez);
  \coordinate (p2) at (\pTwox,\pTwoz);
  \coordinate (p3) at (\pThreex,\pThreez);
  \coordinate (p4) at (\pFourx,\pFourz);
  \draw[line width=6pt,blue] (p1) -- (p3) node[midway,left] {\color{black}$\lambda_{14}$};
  \draw[line width=6pt,blue] (p2) -- (p4) node[midway,right] {\color{black}$\lambda_{32}$};
  \connect{p1}{p3}{bottom color=\antired, top color=red, minimum height=6pt}
  \connect{p4}{p2}{bottom color=red, top color=\antired, minimum height=6pt}
  \shade[ball color = red!100, opacity = 1.] (p1) circle (\radius);
  \shade[ball color = \antired!100, opacity = 1.] (p2) circle (\radius);
  \shade[ball color = \antired!100, opacity = 1.] (p3) circle (\radius);
  \shade[ball color = red!100, opacity = 1.] (p4) circle (\radius);
\end{tikzpicture}  
  \caption[A swing cartoon]{%
    This is a cartoon of the simple swing showcasing the two states that may be
    reconnected to and from.
    Colour connections, here between red--antired charges, are displayed as
    thick lines between pairs of partons.
  }
  \label{fig:simpleswing}
\end{figure}

The general idea is to collect all dipoles between a colour and an
anticolour charge in an event (where a gluon carries both colour and
anticolour charge and thus connects two dipoles together),
consider all possible combinations of two dipoles, and pairwise decide whether
the colour of one dipole instead should be reconnected to form a dipole with the
anticolour of the other and vice versa, as illustrated in \fig{simpleswing}.
A dipole is always assumed to be in a colour singlet state and only dipoles in
the same colour state are permitted to swing.
Technically, this is achieved by assigning a colour index in the range
\mbox{$[0\!:\!8]$} to each dipole and only dipoles with the same index are
allowed to reconnect.
Note that this implies that two dipoles connected by a gluon can never have
the same index since that would correspond to the gluon being in a
colour singlet state.

The imposed requirement that the lambda measure should be minimised means that,
for each potential swing considered, the sum of the lambda measure of the two 
dipoles should be reduced.
In terms of the notation in \fig{simpleswing}, the change
\begin{equation}
  \label{eq:deltalambda}
  \Delta\lambda = \left( \lambda_{14} + \lambda_{32} \right) -
  \left( \lambda_{12} + \lambda_{34} \right)
\end{equation}
should be negative.
In, \eg, \cite{Christiansen:2015yqa}, all possible
such swings are considered and the one with the largest lambda
reduction is performed first.

In the perturbative swing model, additional reconnections that do not reduce
the lambda measure are allowed but they are suppressed by
$e^{-\Delta\lambda}$ and, contrary to most other models where
reconnections are done only once following the parton shower stage, they are
here done continually and interleaved with the final-state parton shower.

The emissions in a typical parton shower are generated in an ordered
sequence using the so-called veto algorithm, \cite{Sjostrand:2006za}, with
the \emph{emission probability} on the form
\begin{equation}
  \label{eq:emprob}
  \mathcal{P}\!\left(p_\perp^2,z\right)
  =
  \frac{\alpha_\mathrm{S}\!\left(p_\perp^2\right)}{2\pi} P_{a\to bc}(z)
  S\!\left(p_\perp^2, p_{\perp\text{prev}}^2\right)
  \mathrm{d}z \frac{\mathrm{d}p_\perp^2}{p_\perp^2},
\end{equation}
where $z$ is the momentum sharing in the splitting 
\mbox{$a\to bc$}
and
\mbox{$P_{a\to bc}(z)$}
is the corresponding Altarelli--Parisi splitting
function. $p_\perp^2$ is the scale of the emission, which in each step
must be smaller than the scale of the previous emission, and
\mbox{$S\big(p_\perp^2, p_{\perp\text{prev}}^2\big)$}
is the probability that there is no emission between the two scales --- the no-emission
factor.\footnote{%
  The no-emission probability is analogous to the Sudakov form factor in QED.
}
In the perturbative swing model, this
probability is supplemented with the probability of performing a swing of two
dipoles:
\begin{equation}
  \label{eq:swingprob}
  \mathcal{P}_\text{sw}\!\left(p_\perp^2\right)
  =
  \swstr e^{-\Delta\lambda}
  S_\text{sw}\!\left(p_\perp^2, p_{\perp\text{prev}}^2\right)
  \frac{\mathrm{d}p_\perp^2}{p_\perp^2},
\end{equation}
where $\swstr$ is an overall swing strength and
\mbox{$S_\text{sw}\big(p_\perp^2, p_{\perp\text{prev}}^2\big)$}
is the
probability that there is no swing between the two scales.

By using the above probability distribution, the swing mechanism will perform a kind of random walk in the
allowed colour configuration space between each emission by the shower.
The explicit aim is that a dipole-based parton shower with continually
inserted reconnections will correctly take
into account the interference between gluons emitted from the coloured
parton and the anticoloured parton of each dipole. One way of looking
at the perturbative swing is that the mechanism approximates the
quadrupole radiation from two colour--anticolour pairs using the weighted
average of the two possible dipole-pair configurations.
The dipole pair-configurations 	of interest are those showcased in the cartoon in \fig{simpleswing}.

Yet another way of looking at the swing mechanism is to recognise that it is
mimicking the full amplitude of a multiparton final state.
We know that for colour
ordered amplitudes of, \eg,
\mbox{e$^+$e$^-\to q\bar{q}+n$g},
the amplitude can be written as a sum over all possible permutations of the
gluons that are allowed by the colour algebra:
\begin{equation}
  \label{eq:colamp}
  \mathcal{A}\propto \sum_{\mbox{perm}[1:n]}\frac{1}{m_{q1}m_{12}m_{23}m_{34}\cdot\ldots\cdot m_{n\bar{q}}}.
\end{equation}
We note that swinging dipoles 12 and 34, with a relative probability
\begin{equation}
  \label{eq:massratio}
  \omega
  =
  \exp\left(-\Delta\lambda\right)
  =
  \exp\big(
    -(\lambda_{14}+\lambda_{32}) + (\lambda_{12}+ \lambda_{34})
  \big)
  =
  \frac{m_{12}^2m_{34}^2}{m_{14}^2m_{32}^2},
\end{equation}
will shuffle the colour state between allowed colour
configurations approximately in proportion to their squared amplitudes and, hence,
introduce effects beyond the
\mbox{$N_c\to\infty$}
approximation of the parton shower.
In order for the random walk-interpretation of this shuffle to hold, it mandates that the relative probability fulfils a \emph{symmetric condition}; symmetric in the sense that when the relative probability to reconnect is proportional to $\omega$, the relative probability to immediately connect back to the original configuration must be proportional to $\omega^{-1}$.
This condition is clearly fulfilled by \eq{massratio}.

\section{A spacetime picture}
\label{sec:spacetime-picture}
So far, we have argued that an intuitive relative probability that may be used
to reconnect a pair of colour dipoles can be obtained by using the dipole
invariant masses in the manner of 
\begin{equation}
  \label{eq:simplew}
  \omega \sim \frac{ s_{12}s_{34} }{ s_{14}s_{32} },
\end{equation}
  where $s_{ij}$ is the invariant mass squared for a colour dipole between partons $i$ and $j$.

  In this section, we provide the justification for the following step: the instances of the squared invariant mass in the weight of \eq{simplew} may naively be replaced by the difference in parton four-space coordinates squared,
\begin{equation}
  \label{eq:modwFac}
  s_{ij} \to {\delta r}_{ij}^2(t).
\end{equation}
It should be pointed out that this clearly is a time-dependent quantity; partons
traverse the collision region and move outwards towards the detector during the
event.
The function
\mbox{${\delta r}_{ij}^2(t)$}
is the time dependent squared spacetime separation of partons $i$ and $j$ and it parametrises the physical extension of a colour dipole spanned by the two partons, arguably more so than the squared invariant mass.
We then aim to relate this time dependence to the evolving
$p_\perp$-scale of the shower mentioned in the previous section.

Based on the relative reconnection probability, as reformulated in terms of
spacetime separation, we have developed and tested a colour reconnection
mechanism intended to describe small collision systems like \ee\ as well as
large systems of colliding Pb nuclei and we describe this mechanism below.

\subsection{The parallel frame, parton vertices, and the proper time}
\label{sec:parallel-frame}
\begin{figure}[t]
  \centering
  \tdplotsetmaincoords{70}{110}
\begin{tikzpicture}[tdplot_main_coords]
  \coordinate (O) at (0,0,0);
  \draw[thick,->] (O) -- (1.8,0,0) node[anchor=north east]{$x$};
  \draw[thick,->] (O) -- (0,1.2,0) node[anchor=north west]{$y$};
  \draw[thick,->] (O) -- (0,0,1.3) node[anchor=south]{$z$};
  \pgfmathsetmacro{\pOnex}{2}
  \pgfmathsetmacro{\pOney}{2}
  \pgfmathsetmacro{\pOnez}{2}
  \pgfmathsetmacro{\pTwox}{-2}
  \pgfmathsetmacro{\pTwoy}{-2}
  \pgfmathsetmacro{\pTwoz}{2}
  \pgfmathsetmacro{\pThreex}{2}
  \pgfmathsetmacro{\pThreey}{-2}
  \pgfmathsetmacro{\pThreez}{-2}
  \pgfmathsetmacro{\pFourx}{-2}
  \pgfmathsetmacro{\pFoury}{2}
  \pgfmathsetmacro{\pFourz}{-2}
  \coordinate (p1) at (\pOnex,\pOney,\pOnez);
  \coordinate (p2) at (\pTwox,\pTwoy,\pTwoz);
  \coordinate (p3) at (\pThreex,\pThreey,\pThreez);
  \coordinate (p4) at (\pFourx,\pFoury,\pFourz);
  \def\connect#1#2#3{%
  \path let
    \p1 = ($(#2)-(#1)$),
    \n1 = {veclen(\p1)},
    \n2 = {atan2(\x1,\y1)} 
  in
    (#1) -- (#2) node[#3, midway, sloped, shading angle=\n2+90, minimum width=\n1, inner sep=0pt, #3] {};
  }
  \def\antired{cyan}
  \connect{p1}{p2}{bottom color=red, top color=\antired, minimum height=6pt}
  \connect{p3}{p4}{bottom color=red, top color=\antired, minimum height=6pt}
  \pgfmathsetmacro{\th}{55}  
  \pgfmathsetmacro{\ph}{45}  
  \pgfmathsetmacro{\eOne}{2.}  
  \pgfmathsetmacro{\eTwo}{2.}  
  \pgfmathsetmacro{\eThree}{2.}  
  \pgfmathsetmacro{\eFour}{2.}  
  \pgfmathsetmacro{\radius}{.2}  
  \pgfmathsetmacro{\radNudge}{1.4}  
  \pgfmathsetmacro{\hOne}{3.4641016}  
  \pgfmathsetmacro{\hTwo}{\hOne}  
  \pgfmathsetmacro{\hThree}{\hOne}  
  \pgfmathsetmacro{\hFour}{\hOne}  
  \pgfmathsetmacro{\vx}{{sin(\th)*cos(\ph)}}
  \pgfmathsetmacro{\vy}{{sin(\th)*sin(\ph)}}
  \pgfmathsetmacro{\vz}{{cos(\th)}}
  \pgfmathsetmacro{\offset}{\radNudge*\radius}
  \tdplotsetrotatedcoordsorigin{(p1)}
  \tdplotsetrotatedcoords{0}{0}{0}
  \draw[dotted,thick,tdplot_rotated_coords] (-\hOne*\vx,-\hOne*\vy,-\hOne*\vz)
    -- (-\offset*\vx,-\offset*\vy,-\offset*\vz);
  \tdplotsetrotatedcoordsorigin{(p2)}
  \tdplotsetrotatedcoords{0}{0}{0}
  \draw[dotted,thick,tdplot_rotated_coords] (\hTwo*\vx,\hTwo*\vy,-\hTwo*\vz)
    -- (\offset*\vx,\offset*\vy,-\offset*\vz);
  \draw[dashed] (O) -- (\pOnex,\pOney,0);
  \draw[dashed] (\pOnex,\pOney,0) -- (\pOnex,\pOney,\pOnez);
  \draw[dashed] (O) -- (\pThreex,\pThreey,0);
  \draw[dashed] (\pThreex,\pThreey,0) -- (\pThreex,\pThreey,\pThreez);
  \tdplotdrawarc[{Triangle[length=2mm]}-]{(O)}{0.7}{-45}{45}{anchor=north west}{$\phi$}
  \tdplotsetthetaplanecoords{45}
  \tdplotdrawarc[-{Triangle[length=2mm]},tdplot_rotated_coords]{(O)}{0.8}{-45}{45}{anchor=south east}{$\theta$}
  \tdplotsetrotatedcoords{20}{90}{90}
  \tdplotsetrotatedcoordsorigin{(p1)}
  \shade[ball color = red!100, opacity = 1.,tdplot_rotated_coords] (p1) circle (\radius);
  \tdplotsetrotatedcoordsorigin{(p2)}
  \shade[ball color = cyan!100,tdplot_rotated_coords] (p2) circle (\radius);
  \tdplotsetrotatedcoordsorigin{(p3)}
  \shade[ball color = red!100,tdplot_rotated_coords] (p3) circle (\radius);
  \tdplotsetrotatedcoordsorigin{(p4)}
  \shade[ball color = cyan!100,tdplot_rotated_coords] (p4) circle (\radius);
  \tdplotsetrotatedcoords{0}{0}{0}
  \tdplotsetrotatedcoordsorigin{(p1)}
  \draw[-{Triangle[length=4mm]},very thick,tdplot_rotated_coords]
    (\offset*\vx,\offset*\vy,\offset*\vz) -- (\eOne*\vx,\eOne*\vy,\eOne*\vz) node[midway,above left] {$q_1$};
  \tdplotsetrotatedcoordsorigin{(p2)}
  \draw[-{Triangle[length=4mm]},very thick,tdplot_rotated_coords]
    (-\offset*\vx,-\offset*\vy,\offset*\vz) -- (-\eTwo*\vx,-\eTwo*\vy,\eTwo*\vz) node[midway,right] {$q_2$};
  \tdplotsetrotatedcoordsorigin{(p3)}
  \draw[-{Triangle[length=4mm]},very thick,tdplot_rotated_coords]
    (\offset*\vx,-\offset*\vy,-\offset*\vz) -- (\eThree*\vx,-\eThree*\vy,-\eThree*\vz) node[midway,above left] {$q_3$};
  \tdplotsetrotatedcoordsorigin{(p4)}
  \draw[-{Triangle[length=4mm]},very thick,tdplot_rotated_coords]
    (-\offset*\vx,\offset*\vy,-\offset*\vz) -- (-\eFour*\vx,\eFour*\vy,-\eFour*\vz) node[midway,above] {$q_4$};
\end{tikzpicture}
  \caption[The parallel frame]{%
    This is an illustration of the parallel frame of reference for the particular case in which all four partons originate from the same spacetime point.
    The momenta are those of \eq{pfmoms} and the partons are colour-connected forming two dipoles.
  }
  \label{fig:pf}
\end{figure}
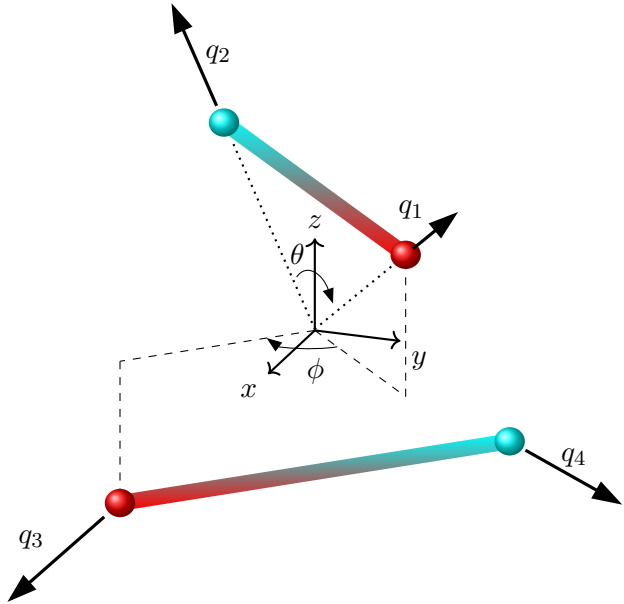

  Any separation in spacetime must necessarily be evaluated in a
  specified frame of reference.
  The \swing\ mechanism joins a suite of
  models already developed for describing collective
  behaviour in terms of string interactions,
  \cite{Bierlich:2014xba, Bierlich:2017vhg}, and a convenient frame of reference
  utilised in these
  models is the \emph{parallel frame}~\cite{Bierlich:2020naj}.
  This frame is, in some sense, the most symmetric frame of reference for
  four partons in terms of their momenta.
  Let partons 1 (colour-charged) and 2 (anticolour-charged) form a colour dipole and
  let partons 3 and 4 do the same.
  The mentioned references have shown
  that for four \emph{massless} partons, the parallel frame can always
  be found (and boosted to from, \eg, the lab frame) such that the
  dipole between partons 1 and 2 is travelling in the $z$-direction,
  while the second dipole between partons 3 and 4 is travelling in the
  opposite direction.
  The parallel frame is completely determined by
  the opening angle, $\theta$, between the momenta of partons 1 and 2
  together with the azimuthal angle, $\phi$, between the projections of the
  two dipoles onto the $xy$-plane, as illustrated in \fig{pf}.
  The momenta, $q_i$, of the four massless partons are given in the
  parallel frame as
\begin{align}
  \label{eq:pfmoms}
  q_1 & = e_1\Big(1;\sin\tfrac{\theta}{2}\cos\tfrac{\phi}{2},\,
                    \sin\tfrac{\theta}{2}\sin\tfrac{\phi}{2},\,
                    \cos\tfrac{\theta}{2}\Big) \nonumber \\
  & \equiv e_1(1; \phantom{-}\mathrm{v}_x, \phantom{-}\mathrm{v}_y, \phantom{-}\mathrm{v}_z), \nonumber \\
  q_2 & \equiv e_2(1; -\mathrm{v}_x, -\mathrm{v}_y, \phantom{-}\mathrm{v}_z), \nonumber \\
  q_3 & \equiv e_3(1; \phantom{-}\mathrm{v}_x, -\mathrm{v}_y, -\mathrm{v}_z), \nonumber \\
  q_4 & \equiv e_4(1; -\mathrm{v}_x, \phantom{-}\mathrm{v}_y, -\mathrm{v}_z),
\end{align}
where $e_i$ is the energy of parton $i$ (\cf Eq.~(4.5)
in~\cite{Bierlich:2020naj}).

\pytppp provides the option to assign production vertices to
partons, \cite{Bierlich:2017vhg}, and we will make extensive use
of this functionality in order to gauge the spatial separation of
partons.

  Using the vertex information, now let the spacetime position of parton $i$, labelled $p_i$, be given as a function of time $t$ according to
\begin{equation}
  \label{eq:pos}
  p_i(t) = v_i + (t - v_{it}) \frac{q_i}{e_i},
\end{equation}
  where $v_i$ is the production vertex of the parton, \ie\ the spacetime point that the parton has been assigned as its production locale, and $v_{it}$ is the temporal coordinate of that point.
  The squared spatial separation between partons $i$ and $j$ can be obtained using the difference in position:
\begin{equation}
  \label{eq:stdSqij}
  {\delta r}_{ij}^2(t) \equiv -\big( p_i(t) - p_j(t) \big)^2.
\end{equation}

  We argue that, given parton production vertices, parton momenta, as well as a temporal evolution variable, the (squared) spatial separation of partons may be calculated for use in a spatially weighted relative probability,
\mbox{$\omega(t)$},
  for a colour reconnection to occur.
  In the introduction to this section it was alluded to how the above spatial separation can be used directly in order to construct such probability weight.
  The justifications of this claim will now be provided using the properties of
  the parallel frame.

  It is straight forward to evaluate the four-vector products needed to obtain the relative probability in \eq{simplew} if the momenta presented in \eq{pfmoms} are used for calculating the invariant masses and the result is
\begin{equation}
  \label{eq:invMassw}
  \frac{ s_{12}s_{34} }{ s_{14}s_{32} }
  =
  \frac{\big( 1 - \mathrm{v}_z^2 \big)^2}{\big( 1 - \mathrm{v}_y^2 \big)^2}.
\end{equation}
  Interestingly, for the analogous ratio of squared spatial separations as
  defined in \eq{stdSqij}, the same result is obtained in the special case where
  all partons originate from the same point, which may be set to
  \mbox{$v_i = 0$} $\forall i$
  without loss of generality, \ie,
\begin{equation}
  \label{eq:sameLim}
  \frac{
    \delta r_{12}^2(t) \delta r_{34}^2(t)
  }{
    \delta r_{14}^2(t) \delta r_{32}^2(t)
  }
  \underset{v_i \to 0}{\to}
  \frac{ s_{12}s_{34} }{ s_{14}s_{32} }.
\end{equation}
  This property motivates further study of the use of the left hand side as a relative probability for colour reconnections taking place within large collision regions.

  Dwelling on the special case where partons originate from the same spacetime vertex, we intend to draw the connection between the spatial separation of partons and the proper time, $\tau$, of the colour dipoles that form between them.
  We make this connection, mainly, because we intend the \swing\ mechanism to be in effect throughout the FSR shower.
  This is to say that each reconnection attempted should compete with the next radiation.
The proper time parameterises the evolution of a dipole allowing it to be
compared with the ordered emissions of the shower.
Additionally, the proper time provides an intuitive measure of when two
dipoles should stop reconnecting and instead enter the hadronisation stage.
  At a certain time, the proper time of an expanding dipole will at some location along the dipole reach a value that we label $\tau_\text{S}$; this is the instance when the colour field has expanded to a maximal girth and the vacuum pressure confines the colour field.
  From this point onwards, the colour field can be treated as a Lund string and the string will hadronise using the built-in \pythia machinery.

  For a Lund string drawn out between two partons that originated at a common spacetime vertex, $v_{ij}$, in the parallel frame, the maximum proper time along the dipole occurs at its center.
To clearly illustrate the connection between the proper time and the spatial
separation (still, we keep the discussion within the parallel frame), consider,
as an example, the momenta of partons 1 and 2 as given in \eq{pfmoms}.
If these partons both originate from the vertex $v_{12}$, then the center point
of the dipole will at time $t$ be at position
\mbox{%
  $(t;x,y,z) = \big(t;v_{12x},v_{12y},v_{12z}+(t-v_{12t})\mathrm{v}_z\big)$.
}
  The proper time at the dipole midpoint, which is the maximum proper time along the dipole, is then
\begin{equation}
  \label{eq:taumax}
  \tau_{12}(t) = (t - v_{12t})\sqrt{1 - \mathrm{v}_z^2}.
\end{equation}
  If this is compared to the physical separation of the dipole ends as given by \eq{stdSqij}, we see in general that
\begin{equation}
  \label{eq:taupropsep}
  \tau_{ij}(t) \propto \sqrt{\delta r_{ij}^2(t)}.
\end{equation}
  Hence, we conclude that the maximum proper time along a colour dipole neatly parameterises the physical separation in the parallel frame of the two partonic dipole ends and we also observe that it is Lorentz invariant in the case when the partons originate from the same spacetime vertex.
  The above relation between physical separation and the maximum proper time will therefore be used to formulate the reconnection probabilities in the \swing\ mechanism proposed in this work.
Note also that the maximum proper time can serve as a proxy for the extent of
transverse field expansion up to a maximal thickness of $\tau_\text{S}$ when
hadronisation begins.

\subsection{Parton separation, dipole vertices, and a new measure}
\label{sec:dipole-separation}
  In most cases, partons in the shower are produced physically close together by the vertex machinery in \pytppp.
The vertex model of this work is no exception to \cite{Bierlich:2017vhg} where
the default vertex assignment smears the production vertex of a parton around
the production vertex of its mother according to a Gaussian distribution for the
offset from the mother's vertex in $x$- and $y$-directions.
The width of this smearing is proportional to the inverse of the $p_\perp$ of
the shower.
A smearing of the production vertex is one way of taking the Heisenberg
uncertainty principle seriously so as to impose that later emissions at lower
$p_\perp$ cannot be precisely resolved.
In the case of \ee\ collisions, coloured partons are instead assigned as their
production vertex the position of their mother at the time of the emission that
produced them.

From the above, we gather that partons that initially form colour dipoles are
expected to be produced close together but, as they traverse the collision
region (and potentially reconnect with partons from other dipoles), the distance
between them and their colour partners will vary according to \eq{stdSqij}.

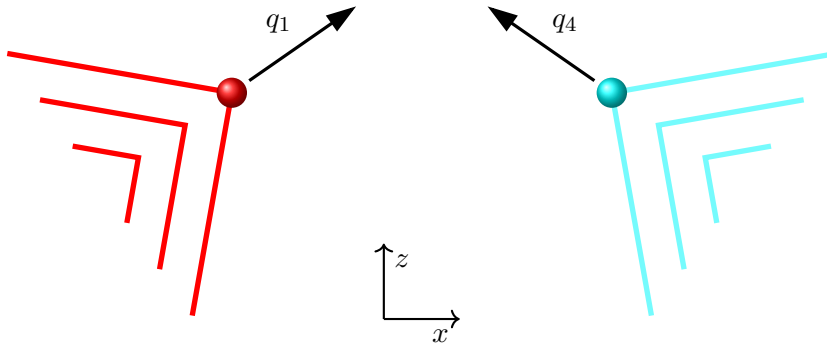
\begin{figure}[t]
  \centering
  \begin{tikzpicture}
  \coordinate (O) at (-0.5,0);
  \draw[thick,->] (O) -- (0.5,0) node[anchor=north east]{$x$};
  \draw[thick,->] (O) -- (-0.5,1) node[anchor=north west]{$z$};
  \pgfmathsetmacro{\pOnex}{-2.5}
  \pgfmathsetmacro{\pOnez}{3}
  \pgfmathsetmacro{\pFourx}{2.5}
  \pgfmathsetmacro{\pFourz}{3}
  \coordinate (p1) at (\pOnex,\pOnez);
  \coordinate (p4) at (\pFourx,\pFourz);
  \pgfmathsetmacro{\th}{55}  
  \pgfmathsetmacro{\eOne}{2.}  
  \pgfmathsetmacro{\eFour}{2.}  
  \pgfmathsetmacro{\radius}{.2}  
  \pgfmathsetmacro{\radNudge}{1.4}  
  \pgfmathsetmacro{\vx}{{sin(\th)}}
  \pgfmathsetmacro{\vz}{{cos(\th)}}
  \pgfmathsetmacro{\offset}{\radNudge*\radius}
  \draw[-{Triangle[length=4mm]},very thick]
    (p1)+(\offset*\vx,\offset*\vz) -- (\pOnex+\eOne*\vx,\pOnez+\eOne*\vz) node[midway,above left] {$q_1$};
  \draw[-{Triangle[length=4mm]},very thick]
    (p4)+(-\offset*\vx,\offset*\vz) -- (\pFourx-\eFour*\vx,\pFourz+\eFour*\vz) node[midway,above right] {$q_4$};
  \pgfmathsetmacro{\len}{3}
  \pgfmathsetmacro{\alpha}{45}
  \pgfmathsetmacro{\sthaOne}{{sin(\th+\alpha)}}
  \pgfmathsetmacro{\cthaOne}{{cos(\th+\alpha)}}
  \pgfmathsetmacro{\sthaTwo}{{sin(\th-\alpha)}}
  \pgfmathsetmacro{\cthaTwo}{{cos(\th-\alpha)}}
  \pgfmathsetmacro{\step}{0.75}
  \def\antired{cyan}
  \definecolor{antilinecolour}{RGB}{117, 251, 253}
  \foreach \n in {0,...,5}
  {
    \pgfmathsetmacro{\scale}{(\len*cos(\alpha) - \n*\step) / (\len*cos(\alpha))}
    \draw[-, color = red, line width=2pt] (\pOnex-\n*\step*\vx-\scale*\len*\sthaOne,\pOnez-\n*\step*\vz-\scale*\len*\cthaOne)
                                   -- (\pOnex-\n*\step*\vx,\pOnez-\n*\step*\vz)
                                   -- (\pOnex-\n*\step*\vx-\scale*\len*\sthaTwo,\pOnez-\n*\step*\vz-\scale*\len*\cthaTwo);
    \draw[-, color = antilinecolour, line width=2pt] (\pFourx+\n*\step*\vx+\scale*\len*\sthaOne,\pFourz-\n*\step*\vz-\scale*\len*\cthaOne)
                                        -- (\pFourx+\n*\step*\vx,\pFourz-\n*\step*\vz)
                                        -- (\pFourx+\n*\step*\vx+\scale*\len*\sthaTwo,\pFourz-\n*\step*\vz-\scale*\len*\cthaTwo);
    \pgfmathparse{(\len*cos(\alpha) - (\n+1)*\step)}
    \ifdim \pgfmathresult pt < 0pt
      \breakforeach
    \fi
  }
  \shade[ball color = red!100, opacity = 1.] (p1) circle (\radius);
  \shade[ball color = \antired!100, opacity = 1.] (p4) circle (\radius);
\end{tikzpicture}
  \caption[Two approaching partons]{%
    This is an illustration of two colour-charged partons approaching each other
    while respectively being part of two separate colour dipoles as indicated by
    colour/anticolour field lines.
    The partons are in general not expected to be in colour contact prior to the
    point of their closest approach.
  }
  \label{fig:approach}
\end{figure}

  The affinity of the parton position as given by \eq{pos} has every parton in the simulated event to trace out a straight world line through the collision region.
The partons are treated as massless, with momenta according to \eq{pfmoms}, and
they as well as the colour field surrounding them propagate with the speed of
light.
Hence, we make the observation that two partons that are approaching each other should not generally have a possibility to strongly interact.
  This point is visualised in \fig{approach} where it is made clear that two approaching dipole ends may not necessarily have come into \emph{colour contact}.
We must also note the issue encountered if we were to reconnect to a dipole
spanned by such approaching partons.
  The behaviour of the colour field for a collapsing dipole is not clear and the usefulness of the otherwise strictly increasing (maximum) proper time along the dipole is lost.

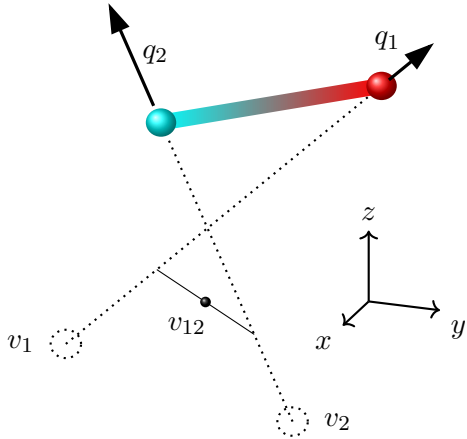
\begin{figure}
  \centering
  \tdplotsetmaincoords{70}{110}
\begin{tikzpicture}[tdplot_main_coords]
  \coordinate (O) at (0,0,0);
  \draw[thick,->] (O) -- (1,0,0) node[anchor=north east]{$x$};
  \draw[thick,->] (O) -- (0,1,0) node[anchor=north west]{$y$};
  \draw[thick,->] (O) -- (0,0,1) node[anchor=south]{$z$};
  \pgfmathsetmacro{\pOnex}{5}
  \pgfmathsetmacro{\pOney}{2}
  \pgfmathsetmacro{\pOnez}{5}
  \pgfmathsetmacro{\pTwox}{-3}
  \pgfmathsetmacro{\pTwoy}{-4}
  \pgfmathsetmacro{\pTwoz}{1}
  \coordinate (p1) at (\pOnex,\pOney,\pOnez);
  \coordinate (p2) at (\pTwox,\pTwoy,\pTwoz);
  \def\connect#1#2#3{%
  \path let
    \p1 = ($(#2)-(#1)$),
    \n1 = {veclen(\p1)},
    \n2 = {atan2(\x1,\y1)} 
  in
    (#1) -- (#2) node[#3, midway, sloped, shading angle=\n2+90, minimum width=\n1, inner sep=0pt, #3] {};
  }
  \def\antired{cyan}
  \connect{p1}{p2}{right color=red, left color=\antired, minimum height=6pt}
  \pgfmathsetmacro{\th}{55}  
  \pgfmathsetmacro{\ph}{45}  
  \pgfmathsetmacro{\eOne}{2.0}  
  \pgfmathsetmacro{\eTwo}{2.0}  
  \pgfmathsetmacro{\radius}{0.2}  
  \pgfmathsetmacro{\radNudge}{1.4}  
  \pgfmathsetmacro{\hOne}{12}  
  \pgfmathsetmacro{\hTwo}{5}  
  \pgfmathsetmacro{\vx}{{sin(\th)*cos(\ph)}}
  \pgfmathsetmacro{\vy}{{sin(\th)*sin(\ph)}}
  \pgfmathsetmacro{\vz}{{cos(\th)}}
  \pgfmathsetmacro{\offset}{\radNudge*\radius}
  \tdplotsetrotatedcoords{0}{0}{0}
  \tdplotsetrotatedcoordsorigin{(p1)}
  \draw[dotted,thick,tdplot_rotated_coords] (-\hOne*\vx,-\hOne*\vy,-\hOne*\vz)
    -- (-\offset*\vx,-\offset*\vy,-\offset*\vz);
  \tdplottransformrotmain{(-\hOne*\vx+\pOnex)}{(-\hOne*\vy+\pOney)}{(-\hOne*\vz+\pOnez)}
  \pgfmathsetmacro{\vOnex}{\tdplotresx}
  \pgfmathsetmacro{\vOney}{\tdplotresy}
  \pgfmathsetmacro{\vOnez}{\tdplotresz}
  \coordinate (v1) at (\vOnex,\vOney,\vOnez);
  \tdplotsetrotatedcoordsorigin{(v1)}
  \tdplotsetrotatedcoords{20}{90}{90}
  \draw[dotted,thick,tdplot_rotated_coords] (v1) circle (\radius);
  \node[left, xshift=-8pt] at (v1) {$v_1$};
  \tdplotsetrotatedcoordsorigin{(p2)}
  \tdplotsetrotatedcoords{0}{0}{0}
  \draw[dotted,thick,tdplot_rotated_coords] (\hTwo*\vx,\hTwo*\vy,-\hTwo*\vz)
    -- (\offset*\vx,\offset*\vy,-\offset*\vz);
  \tdplottransformrotmain{(\hTwo*\vx+\pTwox)}{(\hTwo*\vy+\pTwoy)}{(-\hTwo*\vz+\pTwoz)}
  \pgfmathsetmacro{\vTwox}{\tdplotresx}
  \pgfmathsetmacro{\vTwoy}{\tdplotresy}
  \pgfmathsetmacro{\vTwoz}{\tdplotresz}
  \coordinate (v2) at (\vTwox,\vTwoy,\vTwoz);
  \tdplotsetrotatedcoordsorigin{(v2)}
  \tdplotsetrotatedcoords{20}{90}{90}
  \draw[dotted,thick,tdplot_rotated_coords] (v2) circle (\radius);
  \node[right, xshift=8pt] at (v2) {$v_2$};
  \tdplotsetrotatedcoords{20}{90}{90}
  \tdplotsetrotatedcoordsorigin{(p1)}
  \shade[ball color = red!100, opacity = 1.,tdplot_rotated_coords] (p1) circle (\radius);
  \tdplotsetrotatedcoordsorigin{(p2)}
  \shade[ball color = \antired!100,tdplot_rotated_coords] (p2) circle (\radius);
  \tdplotsetrotatedcoords{0}{0}{0}
  \tdplotsetrotatedcoordsorigin{(p1)}
  \draw[-{Triangle[length=4mm]},very thick,tdplot_rotated_coords]
    (\offset*\vx,\offset*\vy,\offset*\vz) -- (\eOne*\vx,\eOne*\vy,\eOne*\vz) node[midway,above left] {$q_1$};
  \tdplotsetrotatedcoordsorigin{(p2)}
  \draw[-{Triangle[length=4mm]},very thick,tdplot_rotated_coords]
    (-\offset*\vx,-\offset*\vy,\offset*\vz) -- (-\eTwo*\vx,-\eTwo*\vy,\eTwo*\vz) node[midway,right] {$q_2$};
  \pgfmathsetmacro{\vOnet}{0.0}
  \pgfmathsetmacro{\vTwot}{2.0}
  \pgfmathsetmacro{\tclose}{((\vOnet+\vTwot)*(1.0-\vz*\vz) - (\vOnex-\vTwox)*\vx - (\vOney-\vTwoy)*\vy ) / (2.0*(1-\vz*\vz))}
  \coordinate (v1cl) at (\vOnex+\tclose*\vx-\vOnet*\vx,\vOney+\tclose*\vy-\vOnet*\vy,\vOnez+\tclose*\vz-\vOnet*\vz);
  \coordinate (v2cl) at (\vTwox-\tclose*\vx+\vTwot*\vx,\vTwoy-\tclose*\vy+\vTwot*\vy,\vTwoz+\tclose*\vz-\vTwot*\vz);
  \draw (v1cl) -- coordinate (mid) (v2cl);
  \shade[ball color = black] (mid) circle (2pt);
  \node[below left, shift={(4pt,-2pt)}] at (mid) {$v_{12}$};
\end{tikzpicture}
  \caption[Dipole production vertex]{%
    The figure shows two colour-connected partons produced at vertices $v_1$ and $v_2$ with momenta $q_1$ and $q_2$ respectively as seen in the parallel frame of reference.
    The midpoint vertex $v_{12}$ is shown as an illustration of the assigned
    production vertex of the colour dipole.
    This vertex is located at the closest approach in spacetime of the two
    partons.
  }
  \label{fig:dv}
\end{figure}

  Following this line of argumentation, we see that the point of closest approach for two partons is of interest when formulating a colour reconnection mechanism that is competing with the parton shower.
  This vertex of closest approach is illustrated in \fig{dv}.
  It can be computed in the parallel frame through the identification of the time of closest approach, $t^*$, of partons $i$ and $j$.
  Using $t_*$, the partons may then be evolved from their production vertices $v_i$, $v_j$ by making use of \eq{pos} in order to find their respective positions
  \mbox{$p_i(t^*)$} and \mbox{$p_j(t^*)$}.
  The common dipole production vertex, $v_{ij}$, is defined as the midpoint between the partons at $t^*$.

  To find $t^*$, we simply minimise the separation measure in \eq{stdSqij}:
\begin{equation}
  \label{eq:minsep}
  \frac{\mathrm{d}}{\mathrm{d}t} {\delta r}_{ij}^2(t)\bigg\lvert_{t = t^*} = 0,
\end{equation}
  which has the solution
\begin{equation}
  \label{eq:tclose}
  t^*
  =
  \big(
    {\delta v}_{ij} \cdot {\delta u}_{ij}  + (v_{it} + v_{jt}) \, u_i \cdot u_j
  \big)
  \Big/ \abs{\vec{\delta u}_{ij}}^2.
\end{equation}
  Here, the shorthand notation denotes
  \mbox{${\delta v}_{ij} = v_i - v_j$} and
  \mbox{${\delta u}_{ij} = u_i - u_j$} with
  \mbox{$u_i = q_i/e_i$}.

  The average of the parton positions at their closest approach is readily obtainable as
\begin{equation}
  \label{eq:vclose}
  v_{ij} = \frac{p_i(t^*) + p_j(t^*)}{2}.
\end{equation}

Crucially, when the two partonic dipole ends do not originate from the same
spacetime point, this vertex of closest approach will nevertheless be assigned
as the production vertex of the dipole as a way of approximating the origin of
said dipole.
  Such procedure explicitly imprints the quantum mechanical blurring of partonic positions onto the dipole as well.
Most importantly, having a clearly defined position of any dipole makes it
possible to evaluate the spatial separation of dipoles and this is required in
order for us to determine if the colour fields overlap, \ie\ to determine if two
dipoles are close enough in space to interact strongly (reconnect) or not.

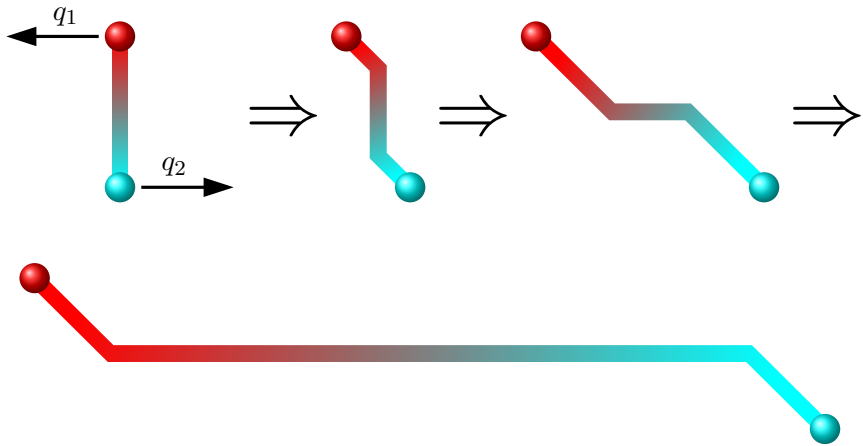
\begin{figure*}[t]
  \centering
  \begin{tikzpicture}
  \pgfmathsetmacro{\pOnex}{0}
  \pgfmathsetmacro{\pOnez}{1}
  \pgfmathsetmacro{\pTwox}{0}
  \pgfmathsetmacro{\pTwoz}{-1}
  \coordinate (p1) at (\pOnex,\pOnez);
  \coordinate (p2) at (\pTwox,\pTwoz);
  \pgfmathsetmacro{\dz}{\pOnez-\pTwoz}
  \def\antired{cyan}
  \def\lw{6pt}
  \shade[top color=red, bottom color=\antired, inner sep=0pt]
         (p1) -- ++(-\lw/2.0,0) -- ++(0,-\dz) -- (p2) -- ++(\lw/2.0,0) -- ++(0,\dz) -- cycle;
  \pgfmathsetmacro{\radius}{.2}  
  \shade[ball color = red!100, opacity = 1.] (p1) circle (\radius);
  \shade[ball color = \antired!100, opacity = 1.] (p2) circle (\radius);
  \pgfmathsetmacro{\eOne}{1.5}  
  \pgfmathsetmacro{\eTwo}{1.5}  
  \pgfmathsetmacro{\radNudge}{1.4}  
  \pgfmathsetmacro{\offset}{\radNudge*\radius}
  \draw[-{Triangle[length=4mm]},very thick]
    (p1)+(-\offset,0) -- (\pOnex-\eOne,\pOnez) node[midway,above,xshift=5pt] {$q_1$};
  \draw[-{Triangle[length=4mm]},very thick]
    (p2)+(\offset,0) -- (\pTwox0+\eTwo,\pTwoz) node[midway,above,xshift=-5pt] {$q_2$};
\end{tikzpicture}
\raisebox{0.9cm}{\resizebox{1cm}{!}{$\Rightarrow$}}
\begin{tikzpicture}
  \def\offset{0.42}
  \pgfmathsetmacro{\root}{1.0/sqrt(2.0)}
  \def\lw{6pt}
  \def\full{\root*\lw/2.0}
  \def\half{\full/2.0}
  \pgfmathsetmacro{\pOnex}{-\offset}
  \pgfmathsetmacro{\pOnez}{1}
  \pgfmathsetmacro{\pTwox}{\offset}
  \pgfmathsetmacro{\pTwoz}{-1}
  \coordinate (p1) at (\pOnex,\pOnez);
  \coordinate (p2) at (\pTwox,\pTwoz);
  \pgfmathsetmacro{\dz}{\pOnez-\pTwoz}
  \def\antired{cyan}
  \pgfmathsetmacro{\ang}{atan2(\pTwox-\pOnex,\pTwoz-\pOnez)}
  \shade[top color=red, bottom color=\antired, shading angle=180-\ang]
         (p1) -- ++(-\full,-\full) -- ++(\offset cm-\half,-\offset cm+\half)
              -- ++(0,-\dz+2*\offset)
              -- ++(\offset cm+\half,-\offset cm-\half) -- ++(\root*\lw,\root*\lw) -- ++(-\offset cm+\half,\offset cm-\half)
              -- ++(0,\dz-2*\offset)
              -- ++(-\offset cm-\half,\offset cm+\half) -- cycle;
  \pgfmathsetmacro{\radius}{.2}  
  \shade[ball color = red!100, opacity = 1.] (p1) circle (\radius);
  \shade[ball color = \antired!100, opacity = 1.] (p2) circle (\radius);
\end{tikzpicture}
\raisebox{0.9cm}{\resizebox{1cm}{!}{$\Rightarrow$}}
\begin{tikzpicture}
  \def\offset{1.0}
  \pgfmathsetmacro{\root}{1.0/sqrt(2.0)}
  \def\lw{6pt}
  \def\full{\root*\lw/2.0}
  \def\half{\full/2.0}
  \pgfmathsetmacro{\pOnex}{-0.5-\offset}
  \pgfmathsetmacro{\pOnez}{1}
  \pgfmathsetmacro{\pTwox}{0.5+\offset}
  \pgfmathsetmacro{\pTwoz}{-1}
  \coordinate (p1) at (\pOnex,\pOnez);
  \coordinate (p2) at (\pTwox,\pTwoz);
  \pgfmathsetmacro{\dx}{\pTwox-\pOnex}
  \def\antired{cyan}
  \pgfmathsetmacro{\ang}{atan2(\pTwox-\pOnex,\pTwoz-\pOnez)}
 \shade[left color=red, right color=\antired, shading angle=180-\ang]
         (p1) -- ++(-\full,-\full) -- ++(\offset cm+\half,-\offset cm-\half)
              -- ++(\dx-2*\offset,0)
              -- ++(\offset cm-\half,-\offset cm+\half) -- ++(\root*\lw,\root*\lw) -- ++(-\offset cm-\half,\offset cm+\half)
              -- ++(-\dx+2*\offset,0)
              -- ++(-\offset cm+\half,\offset cm-\half) -- cycle;
  \pgfmathsetmacro{\radius}{.2}  
  \shade[ball color = red!100, opacity = 1.] (p1) circle (\radius);
  \shade[ball color = \antired!100, opacity = 1.] (p2) circle (\radius);
\end{tikzpicture}
\raisebox{0.9cm}{\resizebox{1cm}{!}{$\Rightarrow$}}
\\\vspace{2em}
\begin{tikzpicture}
  \def\offset{1.0}
  \pgfmathsetmacro{\root}{1.0/sqrt(2.0)}
  \def\lw{6pt}
  \def\full{\root*\lw/2.0}
  \def\half{\full/2.0}
  \pgfmathsetmacro{\pOnex}{-4.2-\offset}
  \pgfmathsetmacro{\pOnez}{1}
  \pgfmathsetmacro{\pTwox}{4.2+\offset}
  \pgfmathsetmacro{\pTwoz}{-1}
  \coordinate (p1) at (\pOnex,\pOnez);
  \coordinate (p2) at (\pTwox,\pTwoz);
  \pgfmathsetmacro{\dx}{\pTwox-\pOnex}
  \def\antired{cyan}
  \pgfmathsetmacro{\ang}{atan2(\pTwox-\pOnex,\pTwoz-\pOnez)}
 \shade[left color=red, right color=\antired, shading angle=180-\ang]
         (p1) -- ++(-\full,-\full) -- ++(\offset cm+\half,-\offset cm-\half)
              -- ++(\dx-2*\offset,0)
              -- ++(\offset cm-\half,-\offset cm+\half) -- ++(\root*\lw,\root*\lw) -- ++(-\offset cm-\half,\offset cm+\half)
              -- ++(-\dx+2*\offset,0)
              -- ++(-\offset cm+\half,\offset cm-\half) -- cycle;
  \pgfmathsetmacro{\radius}{.2}  
  \shade[ball color = red!100, opacity = 1.] (p1) circle (\radius);
  \shade[ball color = \antired!100, opacity = 1.] (p2) circle (\radius);
\end{tikzpicture}
  \caption[A disturbed Lund string]{%
    This is a schematic illustration in the center-of-momentum frame of
    reference of the expansion of a colour dipole where the partonic dipole ends
    originate from nearby but different spacetime vertices.
    The dipole will evolve to become more and more like a Lund string as the
    field expands and becomes confined while the initial field contribution at
    the closest approach of the partons may be neglected late in the evolution.
  }
  \label{fig:wonkyString}
\end{figure*}

  From this point onwards in the discussion, it is important to focus on the fact that partons may not originate from the same spacetime point, either due to the quantum mechanical blurring as simulated by the vertex assignment procedure or due to colour reconnections that connect partons from different regions of the collision volume.
  Partons that are produced with some initial separation in spacetime will never be closer together than at the time $t^*$ that was introduced above in \eq{tclose}.
  It is, however, clear that the maximum proper time along the dipole formed between the partons, see \eq{taumax}, is 0 at this time when the dipole is assumed to be produced at the average vertex defined in \eq{vclose}.
  If the dipole is assumed to evolve from $v_{ij}$ and the partons at the instant corresponding to this positioning are separated by some distance
\mbox{$\sqrt{{\delta r}_{ij}^2(t^*)}$},
  there is some amount of field already present at the time at which the dipole is said to originate.
  This initial separation of dipole ends $i$ and $j$ corresponds to a minimal value,
\mbox{$\tau_{ij0} > 0$},
  of the maximum proper time of the dipole and such observation is in agreement with \eq{taupropsep}, \ie\ that the measure should parameterise the physical separation.
  We therefore modify the measure to be used in the relative probability for a colour reconnection to be a quantity that is better suited to take into account the initial separation of a colour dipole at the closest approach of its partonic ends:
\begin{equation}
  \label{eq:taubar}
  \bar{\tau}_{ij}(t) \equiv \tau_{ij}(t) + \tau_{ij0}.
\end{equation}
  Note about this measure that, in the case where no initial separation of partons is present,
\mbox{($v_i = v_j = v_{ij}$)},
  then
  \mbox{$\bar{\tau}_{ij}^2 \to \tau_{ij}^2 \propto {\delta r}_{ij}^2$}.
The last squared quantity reduces further when the shared vertex of origin is
set to 0,
(\mbox{$v_i = v_j = 0$}),
so that, for our partons that are treated as massless,
  \mbox{${\delta r}_{ij}^2(t) \propto s_{ij} t^2$}
  and the relative probability initially presented in \eq{massratio} for minimising string length is recovered.
  This novel measure reasonably parametrises both the thickness and length of the dipole while (approximately) retaining the useful properties of the proper time that the latter has when used in the traditional sense for a dipole between partons that originate at the same spacetime vertex.
  It is to be assumed that the initial separation as parameterised by $\tau_{ij0}$ may be regarded as ``small" when the colour field has been stretched out to the size of a hadron.
  This assumption is illustrated in the center-of-momentum frame of reference for two partons that span a dipole, see \fig{wonkyString}.
If the initial separation is indeed ``small" later on in the evolution, this
warrants the previous introduced approximation of assigning to the dipole the
production vertex $v_{ij}$.
Any error resulting from the approximation due to some finite minimal
separation between the dipole ends at their closest approach will become less
and less significant during the evolution of the collision system.

  Based on the discussion up to this point, we may now write down a naive expression for a suitable relative probability for colour reconnections that is capable of reducing the length of colour dipoles when evaluated in the parallel frame:
\begin{equation}
  \label{eq:weight}
  \omega(t) \propto \frac{ \bar{\tau}_{12}^2(t)\bar{\tau}_{34}^2(t) }{ \bar{\tau}_{14}^2(t)\bar{\tau}_{32}^2(t) }.
\end{equation}

  Some complications and technicalities remain to be highlighted in the following subsections.

\subsection{Ordering}
\label{sec:ordering}
  The chain of argumentation in the previous sections has lead us to formulate a measure that parameterises the physical extension of a dipole with this measure resembling the proper time of a Lund string and, therefore, it has several convenient properties as outlined in the following.
  The overall objective when introducing the new colour reconnection scheme of this work is to construct a model for colour reconnections that may continuously reconnect colour dipoles that come into vicinity of each other throughout the FSR shower in particle collision events, including everything from \ee\ and \pp\ to \pA\ and \AA\ collisions.
  For every combination of two dipoles that may be considered across the entire event\footnotemark, the aim is to assign a probability for the two dipoles to reconnect at any given scale.
\footnotetext{%
  Fortunately, not all combinations of dipoles have to be considered.
  Based on, \eg, mismatching colour indices or the fact that some dipoles never
  come close enough to each other in order to reasonably have a chance to
  interact during the evolution of the event, some dipole pairs may be discarded
  early on as candidates for a reconnection.
}
  Since colour reconnections modify the colour dipoles involved in the FSR shower, our strategy has been the following:
\begin{enumerate}
  \item Given a current scale in the event generation process (the shower), work
    through all combinations of two dipoles in their parallel frames and
    generate a suggestion for when, \ie\ at some upcoming scale, a reconnection
    happens for any particular dipole pair.
  \item Order the suggestions such that the first colour reconnection to be performed can be selected.
  \item Let this earliest reconnection compete with the upcoming radiation from the shower algorithm. Perform whichever happens at the earlier scale.
  \item Both FSR and colour reconnections modify the dipoles in the event. Hence, repeat from 1. until the hadronisation scale is reached at which point the dipoles have stretched out Lund strings that may be hadronised. Note that the reconnection mechanism can continue past the end of the shower but it stops prior to the onset of hadronisation.
\end{enumerate}

  Here within the suggested scheme lies an issue: how to define \emph{when} during the event generation (\ie\ during the FSR shower) any particular colour reconnection should take place?
The temporal evolution of a pair of dipoles is conveniently addressed in their
parallel frame of reference but it is not possible to compare any time $t$
computed in this frame to corresponding times in other parallel frames for the
purpose of the construction of an ordering of, \eg, reconnections in time.
It is desirable to come up with an invariant scale variable that can be
transferred between parallel frames since this is the frame in which the
calculation of spatial separation is greatly simplified.

  The maximum proper time along any dipole can be used as such evolution variable since it
\begin{itemise}
  \item is Lorentz invariant.
  \item naturally provides a maximum scale\footnote{%
    The proper time $\tau_\text{S}$ is also the point at which, in the Angantyr
    model within \pytppp, Lund strings are assumed to begin to shove each other.
    }
    $\tau_\text{S}$ after which a dipole should no longer
    reconnect but rather hadronise.
  \item is closely connected to the evolution in time $t$ in any given frame of reference.
  \item is, most importantly, directly related to the physical extension of the
    colour dipole. Note that this is only strictly true if the dipole ends
    originate from the same vertex but the concept of the maximum proper time
    may nevertheless still be useful if the measure is constructed to
    incorporate cases when this is nearly true as per the discussion leading up
    to \eq{weight}.
\end{itemise}

  For a temporal-like evolution variable, we will settle on using the maximum proper time along a colour dipole and assume the dipole to originate from an approximated production vertex, see \eq{vclose}.
  The maximum proper time of a dipole, $\tau_{ij}$, is related to the evolution of the dipole in spacetime through its proportionality to time in the parallel frame.
  It may be compared between parallel frames as well as to the inverse shower ordering variable,
\mbox{$1/p_\perp$},
  which is also defined frame independently and which can be regarded as proportional to a proper time of an emission.
The validity of our choice rests on the assumption that the initial separation
$\tau_{ij0}$ loses significance when the colour field expands over time.
Still, dipole pairs that are causally disconnected from other pairs throughout
the collision can not have their respective reconnections strictly temporally
ordered but the proper time treatment does provides an invariant quantity that parameterises the physical extension of the dipoles and which may be compared between frames.

  A technicality arises due to the fact that there are four dipoles of interest when evaluating any given swing and they may all have different values of their respective $\tau_{ij}$ simply due to the fact that they could have been produced at different times which would result in one dipole having expanded for a longer duration than the other three.
  We opt for the choice of ordering the reconnection of a pair according to the \emph{furthest developed} dipole, \ie\ the one with the largest maximum proper time.
The rationale for such choice is that this dipole has a colour field that has
expanded for a longer time compared to the others and will, given that the
initial separation $\tau_{ij0}$ loses significance, be the first dipole to reach
the confinement scale $\tau_\text{S}$ at which point the dipoles should no
longer reconnect; rather, the confined one will hadronise.

Regarding the competition between the colour reconnections and the FSR shower
emissions, a reconnection at a late time should not prevent an emission with a
high $p_\perp$ since the emission would happen earlier.
  Our choice of using the proper time of the furthest developed dipole as a time measure means that the other dipoles may have a much smaller proper time and the reconnection is expected to influence high-$p_\perp$ emissions from such small dipoles.
Clearly, this is an unsolvable dilemma, since the question of \emph{before} or
\emph{after} is one that is frame dependent.
Our choice of ordering means that a reconnection can only affect perturbative
emissions with $1/p_\perp$ larger than any of the proper times of the dipoles
involved.
  Since the parton shower rests on more solid theoretical grounds than our reconnection model, we think that this is a reasonable choice.

\subsection{Restrictions}
\label{sec:restrictions}
  When implementing the colour reconnection concepts discussed above, further aspects that will affect how the spacetime-based \swing\ mechanism can be formulated must be mentioned.

  We made the point in \sec{dipole-separation} that two approaching partons would not in general be expected to have come into colour contact as their colour fields are not necessarily physically overlapping (see \fig{approach}).
However, is it reasonable to demand that all four partons have travelled beyond
their respective closest approach to each other in the parallel frame before
allowing any colour reconnection to take place among them?
We initially indicated that this is the intention and one may, on one hand,
motivate such conviction by considering a colour dipole that is produced, \eg,
by the production of a $q\bar{q}$ pair from a colour singlet.
  In such case, the (anti)colour charges produced and their associated colour field were not present prior to the production of the charge-carrying partons.
  Hence, we argue that such dipoles should not interact before they are actually produced.

  On the other hand, one of the charges carried by a radiated gluon may have existed long prior to the production of the gluon now carrying the charge.
  The field produced by the charge could have had several encounters with other charges before the current dipole end came into existence.

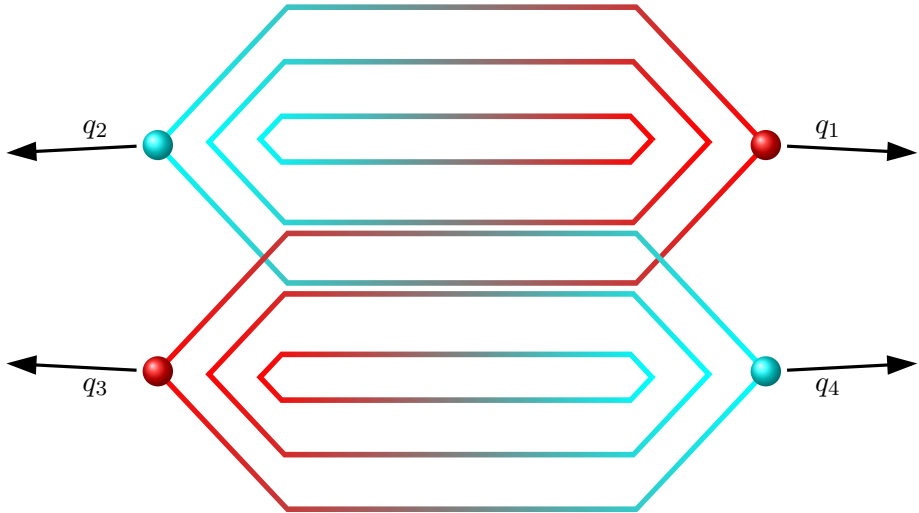
\begin{figure}[t]
  \centering
  \begin{tikzpicture}
  \pgfmathsetmacro{\pOnex}{4}
  \pgfmathsetmacro{\pOnez}{1.5}
  \pgfmathsetmacro{\pTwox}{-4}
  \pgfmathsetmacro{\pTwoz}{1.5}
  \pgfmathsetmacro{\pThreex}{-4}
  \pgfmathsetmacro{\pThreez}{-1.5}
  \pgfmathsetmacro{\pFourx}{4}
  \pgfmathsetmacro{\pFourz}{-1.5}
  \coordinate (p1) at (\pOnex,\pOnez);
  \coordinate (p2) at (\pTwox,\pTwoz);
  \coordinate (p3) at (\pThreex,\pThreez);
  \coordinate (p4) at (\pFourx,\pFourz);
  \pgfmathsetmacro{\th}{93}  
  \pgfmathsetmacro{\eOne}{2.}  
  \pgfmathsetmacro{\eTwo}{2.}  
  \pgfmathsetmacro{\eThree}{2.}  
  \pgfmathsetmacro{\eFour}{2.}  
  \pgfmathsetmacro{\radius}{.2}  
  \pgfmathsetmacro{\radNudge}{1.4}  
  \pgfmathsetmacro{\vx}{{sin(\th)}}
  \pgfmathsetmacro{\vz}{{cos(\th)}}
  \pgfmathsetmacro{\offset}{\radNudge*\radius}
  \draw[-{Triangle[length=4mm]},very thick]
    (p1)+(\offset*\vx,\offset*\vz) -- (\pOnex+\eOne*\vx,\pOnez+\eOne*\vz) node[midway,above left] {$q_1$};
  \draw[-{Triangle[length=4mm]},very thick]
    (p2)+(-\offset*\vx,\offset*\vz) -- (\pTwox-\eTwo*\vx,\pTwoz+\eTwo*\vz) node[midway,above right] {$q_2$};
  \draw[-{Triangle[length=4mm]},very thick]
    (p3)+(-\offset*\vx,-\offset*\vz) -- (\pThreex-\eThree*\vx,\pThreez-\eThree*\vz) node[midway,below right] {$q_3$};
  \draw[-{Triangle[length=4mm]},very thick]
    (p4)+(\offset*\vx,-\offset*\vz) -- (\pFourx+\eFour*\vx,\pFourz-\eFour*\vz) node[midway,below left] {$q_4$};
  \pgfmathsetmacro{\len}{2.5}
  \pgfmathsetmacro{\alpha}{44}
  \pgfmathsetmacro{\sthaOne}{{sin(\th+\alpha)}}
  \pgfmathsetmacro{\cthaOne}{{cos(\th+\alpha)}}
  \pgfmathsetmacro{\sthaTwo}{{sin(\th-\alpha)}}
  \pgfmathsetmacro{\cthaTwo}{{cos(\th-\alpha)}}
  \pgfmathsetmacro{\step}{0.75}
  \def\antired{cyan}
  \foreach \n in {0,...,5}
  {
    \pgfmathsetmacro{\root}{1.0/sqrt(2.0)}
    \def\lw{2pt}
    \def\full{\root*\lw/2.0}
    \def\half{\full/2.0}
    \pgfmathsetmacro{\scale}{(\len*cos(\alpha) - \n*\step) / (\len*cos(\alpha))}
    \coordinate (col1) at (\pOnex-\n*\step*\vx,\pOnez-\n*\step*\vz);
    %
    \shade[right color=red, left color=\antired, shading angle=90]
            (col1) -- ++(\full,-\full) -- ++(\full,\full) --++(-\full,\full)
                   -- ++(-\scale*\len*\sthaOne cm-\half,-\scale*\len*\cthaOne cm+\half)
                   -- ++(-\pOnex cm+\pTwox cm+2*\len*\sthaOne cm-\full,0)
                   -- ++(-\scale*\len*\sthaOne cm-\half,\scale*\len*\cthaOne cm-\half) -- ++(2*\full,-2*\full)
                   -- ++(\scale*\len*\sthaOne cm-\half,-\scale*\len*\cthaOne cm-\half)
                   -- ++ (\pOnex cm-\pTwox cm-2*\len*\sthaOne cm-\full,0)
                   -- ++(\scale*\len*\sthaOne cm-\half,\scale*\len*\cthaOne cm+\half) -- cycle;
    \shade[right color=red, left color=\antired, shading angle=90]
            (col1) -- ++(\full,-\full) -- ++(-\scale*\len*\sthaOne cm-\half,\scale*\len*\cthaOne cm-\half)
                   -- ++(-\pOnex cm+\pTwox cm+2*\len*\sthaOne cm-\full,0)
                   -- ++(-\scale*\len*\sthaOne cm-\half,-\scale*\len*\cthaOne cm+\half)
                   -- ++(-\full,\full) -- ++(2*\full,2*\full) -- ++(\full,-\full)
                   -- ++(\scale*\len*\sthaOne cm-\half,\scale*\len*\cthaOne cm+\half)
                   -- ++ (\pOnex cm-\pTwox cm-2*\len*\sthaOne cm-\full,0)
                   -- ++(\scale*\len*\sthaOne cm-\half,-\scale*\len*\cthaOne cm-\half) -- cycle;
    %
    \coordinate (aco2) at (\pFourx-\n*\step*\vx,\pFourz+\n*\step*\vz);
    \shade[left color=red, right color=\antired, shading angle=90]
            (aco2) -- ++(\full,-\full) -- ++(\full,\full) --++(-\full,\full)
                   -- ++(-\scale*\len*\sthaOne cm-\half,-\scale*\len*\cthaOne cm+\half)
                   -- ++(-\pFourx cm+\pThreex cm+2*\len*\sthaOne cm-\full,0)
                   -- ++(-\scale*\len*\sthaOne cm-\half,\scale*\len*\cthaOne cm-\half) -- ++(2*\full,-2*\full)
                   -- ++(\scale*\len*\sthaOne cm-\half,-\scale*\len*\cthaOne cm-\half)
                   -- ++ (\pFourx cm-\pThreex cm-2*\len*\sthaOne cm-\full,0)
                   -- ++(\scale*\len*\sthaOne cm-\half,\scale*\len*\cthaOne cm+\half) -- cycle;
    \shade[left color=red, right color=\antired, shading angle=90]
            (aco2) -- ++(\full,-\full) -- ++(-\scale*\len*\sthaOne cm-\half,\scale*\len*\cthaOne cm-\half)
                   -- ++(-\pFourx cm+\pThreex cm+2*\len*\sthaOne cm-\full,0)
                   -- ++(-\scale*\len*\sthaOne cm-\half,-\scale*\len*\cthaOne cm+\half)
                   -- ++(-\full,\full) -- ++(2*\full,2*\full) -- ++(\full,-\full)
                   -- ++(\scale*\len*\sthaOne cm-\half,\scale*\len*\cthaOne cm+\half)
                   -- ++ (\pFourx cm-\pThreex cm-2*\len*\sthaOne cm-\full,0)
                   -- ++(\scale*\len*\sthaOne cm-\half,-\scale*\len*\cthaOne cm-\half) -- cycle;
    \pgfmathparse{(\len*cos(\alpha) - (\n+1)*\step)}
    \ifdim \pgfmathresult pt < 0pt
      \breakforeach
    \fi
  }
  \shade[ball color = red!100, opacity = 1.] (p1) circle (\radius);
  \shade[ball color = \antired!100, opacity = 1.] (p2) circle (\radius);
  \shade[ball color = red!100, opacity = 1.] (p3) circle (\radius);
  \shade[ball color = \antired!100, opacity = 1.] (p4) circle (\radius);
\end{tikzpicture}
  \caption[Parallel dipoles]{%
    Two almost parallel dipoles with overlapping colour fields should be able to reconnect even with the dipole ends of the two resulting dipoles approaching.
  }
  \label{fig:paralleldips}
\end{figure}

The necessity of the colour reconnection mechanism to reconnect dipoles before
their endpoints reach their closest approach is most clearly illustrated by the
case of two nearly parallel dipoles of which a cartoon is provided in
\fig{paralleldips}.
The two dipoles showcased have both passed their respective closest approach and
are therefore expanding.
Hence, the behaviour of the field between the parton pairs and the (maximum)
proper time $\tau_{ij}$ (as well as $\bar{\tau}_{ij}$) of the two dipoles are
well defined and the latter is strictly increasing as the colour field expands.
  The dipoles may confidently be said to \emph{exist} at this point in the evolution of the collision and the overlapping fields should interact.
But the two dipoles that would result from reconnecting this system consist of
two pairs of approaching partons for which the behaviour of the colour field is
unclear and the proper time as per \eq{taumax} is not strictly increasing.
  This is particularly problematic if the proper time is to be compared to the evolving (inverse) $p_\perp$ of the parton shower since a decreasing proper time will break the strict ordering of the shower.
  Nevertheless, the dipole configuration illustrated in the figure must be able to interact strongly; the overlapping fields are evident and the two dipoles should therefore be affecting each other.

  The solution proposed in this work is to allow colour reconnections between dipole pairs \emph{iff} either the dipoles of the currently present pair, the \emph{primary} dipoles, both have surpassed their closest approach, \ie\ the two colour fields exist, or, alternatively, the two \emph{secondary} dipoles that will be created by a reconnection both exist.
  This relaxation of the restriction discussed in \sec{dipole-separation} enforces the symmetric condition mentioned in \sec{colo-reconn-pert} on the reconnection model such that two ``shrinking'' dipoles that may have been produced by a reconnection are permitted to connect back to a dipole pair that behaves well.
  This scheme ensures that the involved colour fields are present at the current time even if one of the two dipole configurations contains at least one dipole that does not formally exist (yet).
  If any dipole is formed between two partons that actually share a production vertex (\eg, as is common in \ee\ collisions) it is enforced that such dipole cannot reconnect prior to its production.

  By allowing reconnections that involve dipoles that do not necessarily expand, we must one final time review the relative probability for a reconnection as it was presented in \eq{weight}.
  We argue that the length scale of the closest approach of any dipole as parameterised by $\tau_{ij0}$ limits to this scale the ability to resolve information about the dipole.
  Hence, the measure that parameterises the physical extension of the dipole and provides the reconnection probability, $\bar{\tau}_{ij}$ as seen in \eq{taubar}, should be modified to
\begin{equation}
  \label{eq:maxmod}
  \bar{\tau}_{ij}(t) \to \text{max}\big[ \bar{\tau}_{ij}(t), \tau_{ij0} \big].
\end{equation}

\begin{figure}
  \centering
  \begin{tikzpicture}
  \pgfmathsetmacro{\pOnex}{1.0}
  \pgfmathsetmacro{\pOnez}{-1.0}
  \pgfmathsetmacro{\pTwox}{-1.0}
  \pgfmathsetmacro{\pTwoz}{-1.0}
  \pgfmathsetmacro{\pThreex}{-.7}
  \pgfmathsetmacro{\pThreez}{1.0}
  \pgfmathsetmacro{\pFourx}{.7}
  \pgfmathsetmacro{\pFourz}{1.0}
  \coordinate (p1) at (\pOnex,\pOnez);
  \coordinate (p2) at (\pTwox,\pTwoz);
  \coordinate (p3) at (\pThreex,\pThreez);
  \coordinate (p4) at (\pFourx,\pFourz);
  \def\connect#1#2#3{%
  \path let
    \p1 = ($(#2)-(#1)$),
    \n1 = {veclen(\p1)},
    \n2 = {atan2(\x1,\y1)} 
  in
    (#1) -- (#2) node[#3, midway, sloped, shading angle=\n2+90, minimum width=\n1, inner sep=0pt, #3] {};
  }
  \def\antired{cyan}
  \connect{p1}{p2}{right color=red, left color=\antired, minimum height=6pt}
  \connect{p3}{p4}{left color=red, right color=\antired, minimum height=6pt}
  \pgfmathsetmacro{\th}{35}  
  \pgfmathsetmacro{\eOne}{1.}  
  \pgfmathsetmacro{\eTwo}{1.}  
  \pgfmathsetmacro{\eThree}{1.}  
  \pgfmathsetmacro{\eFour}{1.}  
  \pgfmathsetmacro{\radius}{.2}  
  \pgfmathsetmacro{\radNudge}{1.4}  
  \pgfmathsetmacro{\offset}{\radNudge*\radius} 
  \pgfmathsetmacro{\hOne}{abs(\pOnex-\pTwox)/2.0/sin(\th)}  
  \pgfmathsetmacro{\hTwo}{\hOne}  
  \pgfmathsetmacro{\hThree}{abs(\pThreex-\pFourx)/2.0/sin(\th)}  
  \pgfmathsetmacro{\hFour}{\hThree}  
  \pgfmathsetmacro{\vx}{{sin(\th)}}
  \pgfmathsetmacro{\vz}{{cos(\th)}}
  \coordinate (O) at (-5,\pOnez-\hOne*\vz);
  \draw[thick,->] (O) -- (-4,\pOnez-\hOne*\vz) node[anchor=north east]{$x$};
  \draw[thick,->] (O) -- (-5,\pOnez-\hOne*\vz+1) node[anchor=north west]{$z$};
  \draw[dotted,thick] (\pOnex-\hOne*\vx,\pOnez-\hOne*\vz)
    -- (\pOnex-\offset*\vx,\pOnez-\offset*\vz);
  \draw[dotted,thick] (\pTwox+\hTwo*\vx,\pTwoz-\hTwo*\vz)
    -- (\pTwox+\offset*\vx,\pTwoz-\offset*\vz);
  \draw[dotted,thick] (\pThreex+\hThree*\vx,\pThreez+\hThree*\vz)
    -- (\pThreex+\offset*\vx,\pThreez+\offset*\vz);
  \draw[dotted,thick] (\pFourx-\hFour*\vx,\pFourz+\hFour*\vz)
    -- (\pFourx-\offset*\vx,\pFourz+\offset*\vz);
  \node[below right] at (\pOnex-\hOne*\vx,\pOnez-\hOne*\vz) {$v_{12}$};
  \node[above right] at (\pThreex+\hThree*\vx,\pThreez+\hThree*\vz) {$v_{34}$};
  \shade[ball color = red!100, opacity = 1.] (p1) circle (\radius);
  \shade[ball color = \antired!100, opacity = 1.] (p2) circle (\radius);
  \shade[ball color = red!100, opacity = 1.] (p3) circle (\radius);
  \shade[ball color = \antired!100, opacity = 1.] (p4) circle (\radius);
  \draw[-{Triangle[length=4mm]},very thick]
    (p1)+(\offset*\vx,\offset*\vz) -- (\pOnex+\eOne*\vx,\pOnez+\eOne*\vz) node[midway,below right] {$q_1$};
  \draw[-{Triangle[length=4mm]},very thick]
    (p2)+(-\offset*\vx,\offset*\vz) -- (\pTwox-\eTwo*\vx,\pTwoz+\eTwo*\vz) node[midway,below left] {$q_2$};
  \draw[-{Triangle[length=4mm]},very thick]
    (p3)+(-\offset*\vx,-\offset*\vz) -- (\pThreex-\eThree*\vx,\pThreez-\eThree*\vz) node[midway,above left] {$q_3$};
  \draw[-{Triangle[length=4mm]},very thick]
    (p4)+(\offset*\vx,-\offset*\vz) -- (\pFourx+\eFour*\vx,\pFourz-\eFour*\vz) node[midway,above right] {$q_4$};
  \pgfmathsetmacro{\side}{2.2}
  \draw[-|,thick] (\side,\pOnez) -- (\side,\pOnez-\hOne*\vz) node[midway,right]{$(t-v_{12t})\mathrm{v}_z$};
  \draw[-|,thick] (\side,\pFourz) -- (\side,\pFourz+\hFour*\vz) node[midway,right]{$(t-v_{34t})\mathrm{v}_z$};
  \draw[|-|,thick] (\side,\pOnez) -- (\side,\pFourz) node[midway,right]{$\left|\Delta(t)\right|$};
  \draw[|-|,thick] (-\side,\pOnez-\hOne*\vz) -- (-\side,\pFourz+\hFour*\vz) node[midway,left]{$\left|d_0\right|$};
\end{tikzpicture}
  \caption[Colour field overlap]{%
    This is a cartoon of two colour dipoles approaching each other in the
    $z$-direction.
    The dipoles are assumed to be produced at vertices $v_{12}$ and $v_{34}$ as
    seen in the parallel frame of reference.
    The separation between the two dipoles are marked in the figure with
    \mbox{$\Delta(t) = (2t - v_{12t} - v_{34t}) \mathrm{v}_z + d_0$}.
  }
  \label{fig:cov}
\end{figure}
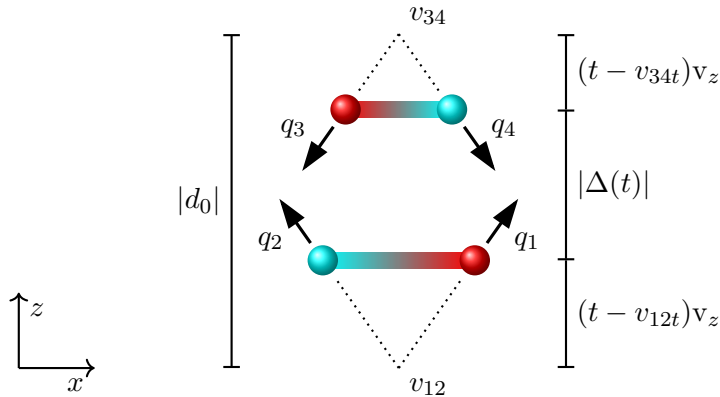

  Emphasis should be put on the fact that dipoles are only allowed to reconnect if the colour fields are overlapping, \ie\ there is colour contact between the dipoles.
  This results in a further restriction in our implementation.
  In practice, the restriction entails that a pair of dipoles may only reconnect if either the two primary dipole fields are overlapping, the two secondary dipole fields are overlapping, or if all four dipole fields are overlapping.
  This deals with cases where two dipoles may be produced well separated in space but as they traverse the collision region the colour fields may begin to overlap and one or several reconnections could take place before the dipoles pass each other and the fields no longer overlap.
  A crude model for determining if two dipoles are overlapping has been implemented.
  The method consists of a simple comparison between the physical separation of the dipole center points to the thickness of the furthest expanded colour field, the latter parameterised by \mbox{$\tau_{ij}(t)$}.
The former separation is the spatial separation only along the axis of direction
of travel in the parallel frame.
  A cartoon of a case of two approaching dipoles is provided in \fig{cov}.
  For the example of the separation between dipole 12 and 34, the dipole separation is given as
\begin{equation}
  \label{eq:separation}
  \Delta(t) = (2t - v_{12t} - v_{34t}) \mathrm{v}_z + d_0,
\end{equation}
where
\mbox{$d_0 = v_{12z} - v_{34z}$}
and it is the separation in the $z$-direction of the two assigned dipole
production vertices.

  As stated earlier, dipoles are not allowed to reconnect if the proper time of any of the dipoles at the instance of reconnection exceeds $\tau_\text{S}$.

  One additional restriction is imposed on the secondary dipoles.
The reconnection algorithm employs Monte Carlo techniques with the consequence
that a reconnection will not strictly minimise the length of colour dipoles in
each instant of a reconnection taking place but only on the average over
multiple reconnections.
In order to prevent large dipoles from forming, \ie\ secondary dipoles with an
initial separation
\mbox{$\tau_{ij0} > \tau_\text{S}$},
  any reconnection where one or two such large secondary dipoles would be produced is simply vetoed.
  The partons of such dipoles never come close enough to form a dipole smaller than the confinement scale.

\subsection{The \swing\ strength}
\label{sec:swing-strength}
Following the above discussion of the colour reconnection model we supply,
including the restrictions imposed as of \sec{restrictions}, it is now in place
to dissect the resulting relative probability which we will label the
\emph{weight function}.
  It is a function of the time $t$ in the parallel frame and comes on the form
\begin{equation}
  \label{eq:finalweight}
  \omega(t)
  \propto
  \frac{
    \text{max}\big[ \bar{\tau}_{12}(t), \tau_{12,0} \big]^2
    \text{max}\big[ \bar{\tau}_{34}(t), \tau_{34,0} \big]^2
  }{
    \text{max}\big[ \bar{\tau}_{14}(t), \tau_{14,0} \big]^2
    \text{max}\big[ \bar{\tau}_{32}(t), \tau_{32,0} \big]^2
  }.
\end{equation}
We note that it is the product of two second order polynomials in $t$ divided by
two other second order polynomials in $t$ unless one or more of the max-value
functions reduce the order of any polynomial.
This weight function may change its behaviour at several instances over the
evolution in time, in particular at any of the max-value function breakpoints
$\tau_{ij0}$, which complicates its use as a relative probability for colour
reconnections.
  The proportionality constant indicated in the above expression has been introduced as a parameter in our code in analogy with the swing strength $\alpha_\text{sw}$ in \eq{swingprob}.

The weight function will always be applied within a restricted time interval
since a pair of dipoles may not reconnect prior to the time at which either the two primary or secondary dipoles have evolved past their closest approach; this provides a minimum time at which the dipole pair may reconnect.
Further, if any of the four possible dipoles of the pair configuration evolves
up to a time at which its proper time becomes equal to $\tau_\text{S}$, this
time is the maximum time at which the dipole pair may reconnect.
On the interval between the minimum and maximum values, the weight function may
be considered piecewise and divided up into as many subintervals as needed, each
on which the function is strictly increasing or decreasing.
Any maxima or minima of the weight function located in between any of the
max-value function break points $\tau_{ij0}$ are easily identifiable in each of
the cases of different polynomial order of the numerator and denominator.

  The restrictions discussed in the previous section implies a technicality which is carried over to the analysis of the weight function.
  If two partons that form any of the secondary dipoles are produced at the same vertex, the weight function diverges at this production time (separation, and therefore the proper time, is 0).
However, there is no reason as to why such dipole cannot be interacting from the
very moment it is produced and the algorithm must be capable of identify the
divergens and permit a reconnection from this very point and onwards.
Consider as a relevant example that the two primary dipoles may both exist and
overlap and, so, they may reconnect at some time $t$ where they instead form the
secondary dipoles.
Any of the secondary dipoles, however, may cause a divergens if the partons of
any secondary dipole originate from a single point.
This point of origin may happen after, \ie\ at a later time than, the computed reconnection time.
  We identify that the restriction criteria may be fulfilled by the primaries with the result that a reconnection could be tried before the closest approach of the secondaries and the later may cause a divergens.
  Since it has been argued that such a secondary dipole cannot be allowed to reconnect prior to its production, the reconnection is instead forced to happen at the divergence point (later in time compared to the suggested reconnection).

  The final point to be discussed is the possibility of a \emph{dead} interval in time.
Since we require either the primary or the secondary dipoles to physically
overlap, there is a possibility of a gap between the periods of overlap.
This is handled by not allowing reconnections in such time windows but if any of
the dipole pairs come into colour contact later on in time, they are again
permitted to reconnect.

\subsection{The final algorithm}
\label{sec:final-algorithm}
This section provides a summary and brief overview of our implemented \swing\
algorithm; a detailed overview together with technical notes has been included
in \app{details}.
We note that the procedure described here, as well as the results presented
further down, are specific to the current version of our code.
This code has not been published as it is part of the ongoing development of
the \pythia program.
Nonetheless, interested readers are welcome to contact us with any questions or
requests and we will happily provide the code used to produce the results in
this paper.

We have implemented a colour reconnection mechanism that continually competes
with the FSR shower by determining whether to let the shower radiate a parton or
to reconnect two colour dipoles.
The preparation of dipoles for the \swing\ proceeds as follows:
\begin{enumerate}
  \item The \swing\ mechanism considers all pairs of dipoles with matching colour states, as a match is required for a reconnection to take place.
    Dipoles that share a parton at one of their ends are prohibited from reconnecting as such reconnections would form colour singlets.
  \item For each dipole pair, the parallel frame is computed and, if successfully determined, stored for reuse.
    If the parallel frame cannot be computed---such as when the dipole invariant mass is very small---the dipoles are considered unable to reconnect.
  \item The different restrictions detailed in \sec{restrictions}, are
    taken into account to determine the interval(s) in $\tau$ where
    reconnections are allowed.
  \item An attempt is made to generate a reconnection scale $\tau$ for the
    dipole pair.
    If this attempt is successful, the reconnection scale is stored for the
    specific dipole pair.
    This scale will be compared to the inverse $p_\perp$-scale of the shower.
\end{enumerate}

The above initialisation of the \swing\ identifies a subset of all dipole pairs;
those that are permitted to reconnect and, for each pair, at which scale they
may do so provided that nothing happens at an earlier scale.
  The \swing\ mechanism can then compete with the FSR shower and potentially reconnect multiple dipole pairs between any two radiated partons.
  When a parton is radiated by the shower or when a colour reconnection takes place, a subset of the dipoles present in the collision system is changed.
  The algorithm therefore keeps track of which dipole(s) are affected by either the \swing\ mechanism or the shower and reruns the preparation steps described above for any affected or newly created dipoles.
  In addition, the shower algorithm is informed about any newly created or changed dipoles. 
  This procedure is repeated throughout the evolution of the shower.

\section{Results}
\label{sec:res}
  %
Simulated event multiplicities together with other selected event
properties obtained from simulations that include our new \swing\
mechanism are presented in this section with comparisons to referenced
experimental data.

The introduction of an alternative colour reconnection model necessitates tuning
it to data.
Values of new parameters introduced in the model as well as values of parameters
that are already part of the \pythia program should be considered for
the tuning.
Here, we follow the same procedure as in \cite{Lonnblad:2023stc}, where a simplified modification of the QCD-based reconnection model was applied in heavy-ion collision simulations.
The modifications are implemented as follows:
  adjust the parameters of the new model and relevant parameters in \pythia to give a good description of minimum-bias \pp\
data;
  then adjust parameters of the Angantyr model in \pythia to obtain also a good description of p$A$ data;
  and, finally, extrapolate this to \AA\ data for which no additional parameters are available for
adjustments.

As discussed above, our new \swing\ model is just one part of a larger
framework where dense collision systems are described in terms on
interactions between strings. We have therefore
opted not to do a sophisticated full tuning of the Gleipnir code, but
instead focused on modifying the values of a selection of a few
parameters that directly influence particle multiplicities in
collisions so as to get an acceptable description of available data.

Our \swing\ code introduces two parameters: the swing strength,
\swstr, that controls how often we try to swing and a parameter
that controls the cut on large secondary dipoles, see
\sec{restrictions}.  However, we mention here that when varying the
values of these parameters around unity, no significant impact on the
simulation results investigated was found.  Regarding specifically the
swing strength parameter, only when its value was reduced by an order
of magnitude are significant effects achieved and these effects are,
in line with expectations, consistent with constraining the
possibility of reconnections towards a point at which reconnections do
not occur at all.  The seemingly appropriate unit value of the swing
strength parameter should be understood in terms of saturation; an
increased swing strength does indeed make reconnections more likely to
happen, as was verified, but, regardless of the increase in the number
of reconnections taking place, the reconnection mechanism does not
present to the shower a significantly different sample of dipole
configurations to radiate from compared to what the default strength
does.  Akin to a thermalised system, the number of reconnections
performed with the default strength is sufficient for the algorithm to
effectively explore the energy landscape of the colour dipoles and to find the
minimum energy of the system in between each shower splitting.

In the following \sec{eeres}--\ref{sec:PbPbres}, the results of our
rudimental tuning process and the simulation results that ensued from
it will be provided for each investigated collision system in turn.
In addition to the procedure in \cite{Lonnblad:2023stc}, we here begin with
studying \ee\ collisions, before addressing \pp, pPb, and, finally, PbPb
collisions.

The analyses are displayed and compared to experimental data with the
help of the Rivet toolkit (version \mbox{3.1.6}) and, when relevant,
its support for analysis of simulated heavy-ion collision
data~\cite{Bierlich:2019rhm, Bierlich:2020wms}.

\subsection{\texorpdfstring{\ee\ results}{e+e- results}}
\label{sec:eeres}
The original perturbative \swing\ model was shown to give some effects
in \ee\ annihilations at LEP~\cite{Lonnblad:1995yk}.
Such effects are naively expected to be small since, in the parton shower, one starts out with only one dipole stretching between the $q$ and the $\bar{q}$ from the Z$^0$ decay.
It is only after two gluon emissions in the partonic
cascade that we have a possibility of finding two dipoles in the same
colour state which then may reconnect or not.

In our new colour reconnection mechanism, the spacetime weight function (\eq{finalweight}) is constructed to reproduce the momentum-based function in the original perturbative swing model in the limit where all spacetime differences in the production points of partons are small.
Since this can be assumed to be the case in Z boson decays in \ee\ collisions,
is a good process to use an initial validation of our model.

\begin{figure*}[t]
  \centering
  \includegraphics[width=0.48\textwidth]{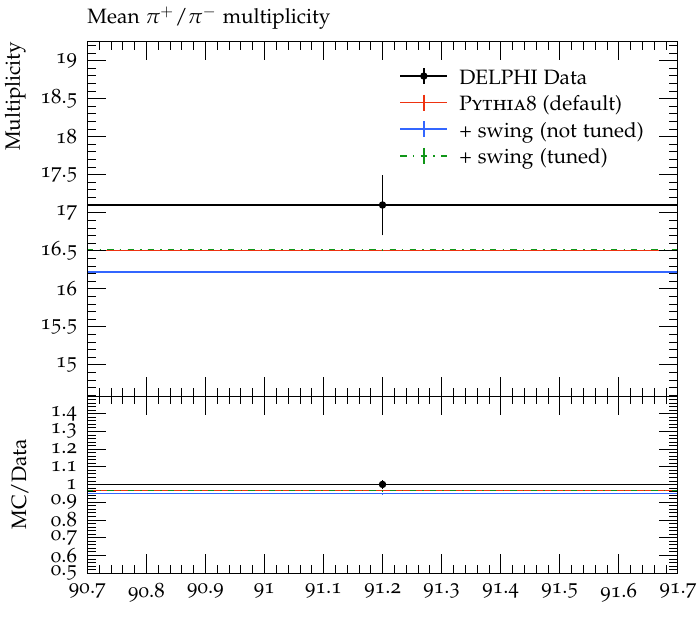}
  \hfill
  \includegraphics[width=0.48\textwidth]{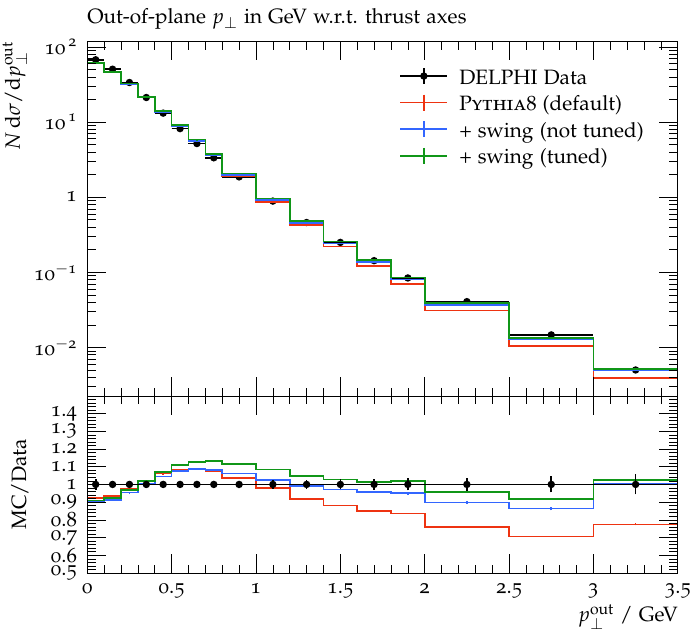}
  \caption[\ee\ event shape results]{%
    This plot compares the tuned \swing\ mechanism, a version not tuned, as well
    as the default \pythia to \mbox{\ee} data from LEP~\cite{DELPHI:1996sen}.
    The comparison motivates the choice of retuning the $\alpha_\text{S}$
    parameter of the FSR shower from the default of 0.1365 to 0.14.
    The effects of the (tuned) \swing\ mechanism on \ee\ for charged pion mean
    multiplicities (left) as well as event shapes in $p_\perp$ (right) are
    reassuring when compared to \pythia 's default settings as well as to the
    experimental data~\cite{DELPHI:1996sen}.
    We note that in the multiplicity plot, the tuned swing mechanism in the
    dash-dotted line coincides with the default \pythia curve.
    \ee\ events were generated at
    \mbox{$\sqrt{s} = 91.2$~GeV}.
  }
  \label{fig:eeres}
\end{figure*}

The set of commands and parameter values used for the simulation setup
is provided in \app{ee-setup}.
When comparing the overall multiplicity as generated by the \pythia program with
and with out the new \swing, we find, as expected, that the \swing\ gives a
slight reduction.
The effect is shown in \fig{eeres} (left) where the average $\pi^\pm$
multiplicity is compared to DELPHI data~\cite{DELPHI:1996sen}.
Also shown is the result of simulations with the \swing\ after tuning slightly the reference value for
\mbox{$\alpha_\text{S}(M_\text{Z}^2)$},
used in the \pythia FSR shower, from the default value of \mbox{0.1365} to \mbox{0.14}.
With this new
value, most event shapes and multiplicities are either restored to the default behaviour of \pythia\ or improved upon and, therefore, this is the value that we
will use for the final-state parton shower parameter also when applying the new
\swing\ to hadronic collisions further down this section.

An important aspect to note is that there are some event shapes for which the effects of the \swing\ cannot be tuned away.
In particular, we show in \fig{eeres} (right) the effect on the transverse momentum distribution out of the plane defined by the thrust axis and the thrust-major axis in each event.
On a perturbative level, the leading order contribution to this observable is $\alpha_\text{S}^2$ and, as mentioned above, it is only after the second gluon emission that we expect to see an effect of the perturbative swing.
As seen in the figure, the effect is still there
also after the tuning and is even slightly enhanced.
This is exactly the effect that was seen in \cite{Lonnblad:1995yk} for the
original momentum-based perturbative swing model and we therefore feel
confident that our new model is validated in this respect.

\subsection{\pp\ results}
\label{sec:ppres}
Much of the development work on our colour reconnection model focused
on \pp\ collisions for which simulations were performed and event
properties were studied.  The set of commands and parameter values
used for the simulation setup is provided in \app{pp-setup}.

Contrary to the description of \ee\ collisions, simulations of \pp\
events that invoke the \swing\ mechanism must by design utilise the
heavy-ion machinery of \pytppp.  A major reason for this choice was
the desire to have a common framework for handling FSR and colour
reconnections globally across the entire event regardless of the
number of hadrons involved in a collision.  This would give a
consistent treatment of collision systems ranging in size from \pp\ to
\AA.

The default behaviour of the heavy-ion module is to generate the full
partonic state of each nucleon-nucleon (NN) subevent separately prior to
stacking them together and hadronising the full partonic state.
Like the hadronisation procedure, however, we want our new colour reconnection
mechanism to act globally on all the colour dipoles.
The realisation of such global scheme has demanded the explicit separation of
FSR from the process of generating initial-state radiation (ISR) and MPI for
each separate subevent.

\begin{figure}[t]
  \centering
  \includegraphics[width=0.48\textwidth]{%
    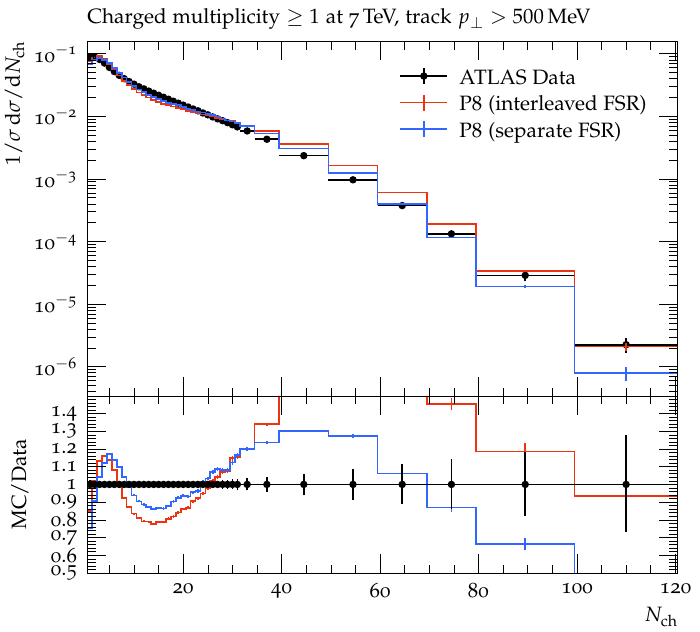}
  \caption[\pp\ globalFSR]{%
    This figure illustrates the difference caused by separating out the FSR from
    the ISR and MPI rather that generating FSR, ISR, and MPI fully interleaved
    as is the default in \pythia.
    Minimum bias \pp\ events were generated using Angantyr at
    \mbox{$\sqrt{s} = 7$~TeV}
    without invoking the \swing\ mechanism.
    The effects due to the separation of FSR must be tuned away when the full
    \swing\ is introduced.
    \pp\ data from ATLAS, \cite{ATLAS:2010jvh}, has been included for reference.
  }
  \label{fig:globalFSR}
\end{figure}

Technically, this is easy to do, but as the FSR in the default
\pythia setup is \textit{interleaved} with the MPI and ISR generation,
\cite{Sjostrand:2004ef}, the end result of simulations will differ when the FSR
is separated out compared to when it is interleaved.
The fact that the FSR in the interleaved case is competing with the MPI and ISR
emissions means that the phase space available for FSR, when instead separately performed after the MPI+ISR, will be different for the separated FSR.
As a consequence, even in the case of \pp\ collisions where there is only one
subevent, the act of separating the FSR gives a noticeable effect as is
illustrated in \fig{globalFSR}.

When we include our new \swing\ mechanism in the generation of \pp\
collisions, we need to switch off the default colour reconnections as they are normally
done in \pythia after the MPI+ISR(+FSR) step and the net result is
that multiplicities in general increases indicating that our model
results less reconnections and hence longer strings than the default
setup.
Before we jump to such conclusions, however, we know that other
things also are changed, \eg, the radiation pattern, and we therefore need to retune parameters in
\pythia.

\begin{figure*}[t]
  \centering
  \includegraphics[width=0.48\textwidth]{%
    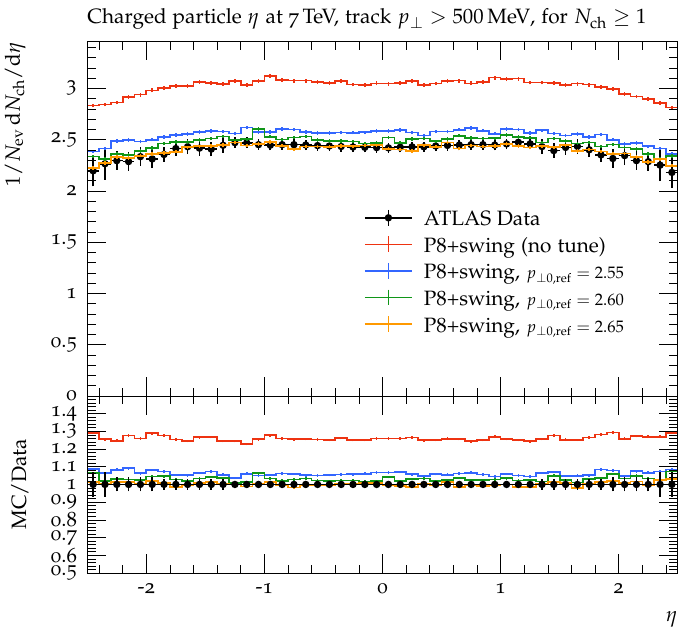}
  \hfill
  \includegraphics[width=0.48\textwidth]{%
    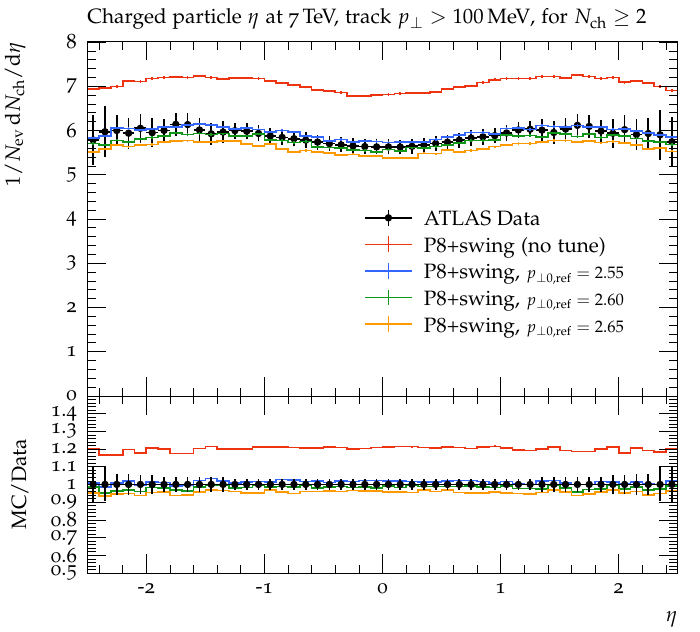}
  \caption[\pp\ tune]{%
    This plot of charged-particle multiplicities in pseudorapidity $\eta$ for
    \pp\ events generated at
    \mbox{$\sqrt{s} = 7$~TeV}
    compared to minimum bias ATLAS \pp\ data, \cite{ATLAS:2010jvh}, motivates
    the choice of retuning the parameter $p_{\perp0,\text{ref}}$ from the
    default 2.28~GeV to 2.60~GeV.
    Note the different cuts for each plot.
  }
  \label{fig:pptune}
\end{figure*}

A key parameter that influences the overall multiplicity is the
$p_{\perp0,\text{ref}}$ scale in the formula for the energy dependent
suppression parameter $p_{\perp0}$ in the MPI and ISR and we have
chosen to perform a rudimentary returning of this $p_{\perp0,\text{ref}}$.
Some results of the retuning procedure are illustrated in \fig{pptune}
where the effect on the pseudorapidity
distribution of charged particles (with two different cuts on the
minimum transverse momentum and minimum number of particles) for
different $p_{\perp0,\text{ref}}$ settings are shown for \pp\ collisions at
\mbox{$\sqrt{s}=7$~GeV}.
It turns out that the value of
$p_{\perp0,\text{ref}}$ needed to get reasonable multiplicities is
around 2.6~GeV and this is quite a bit higher than the 2.28~GeV used by
the default \pytppp setup.

\begin{figure*}[!t]
  \centering
  \begin{minipage}{0.48\textwidth}
    \centering
    \includegraphics[width=\textwidth]{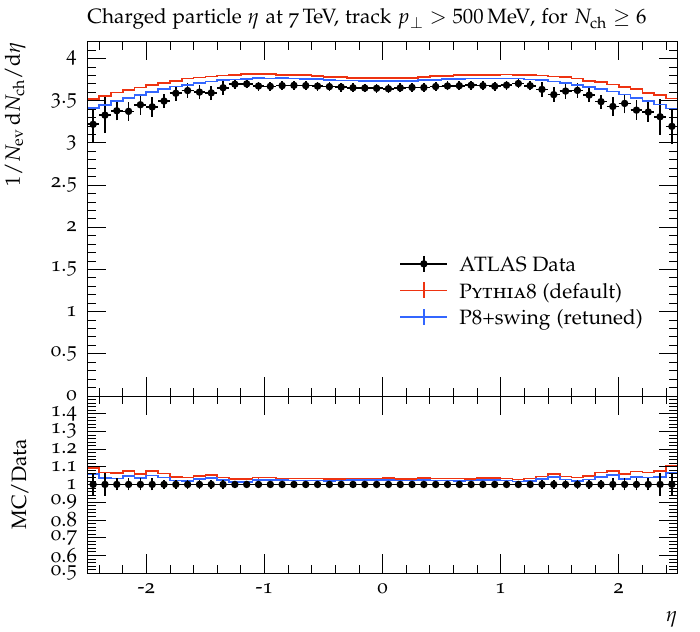}
  \end{minipage}%
  \hfill
  \begin{minipage}{0.48\textwidth}
    \centering
    \includegraphics[width=\textwidth]{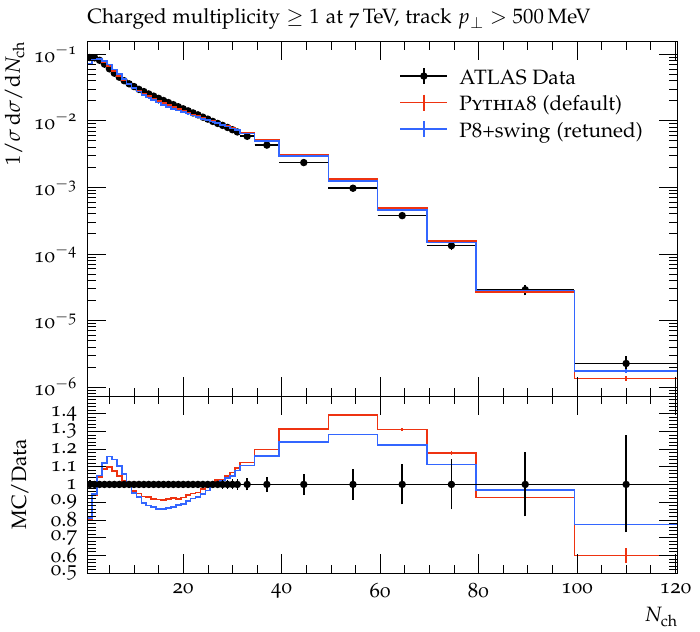}
  \end{minipage}%
  \vspace{1em}
  %
  \begin{minipage}{0.48\textwidth}
    \centering
    \includegraphics[width=\textwidth]{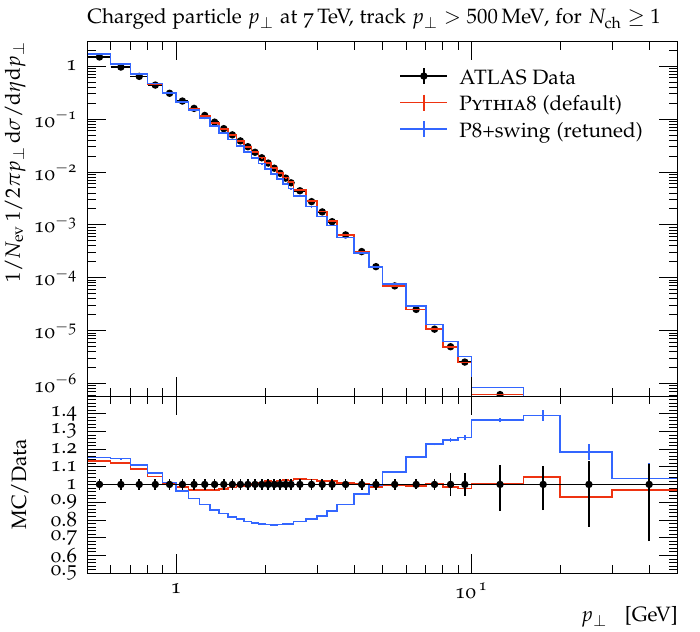}
  \end{minipage}%
  \hfill
  \begin{minipage}{0.48\textwidth}
    \centering
    \includegraphics[width=\textwidth]{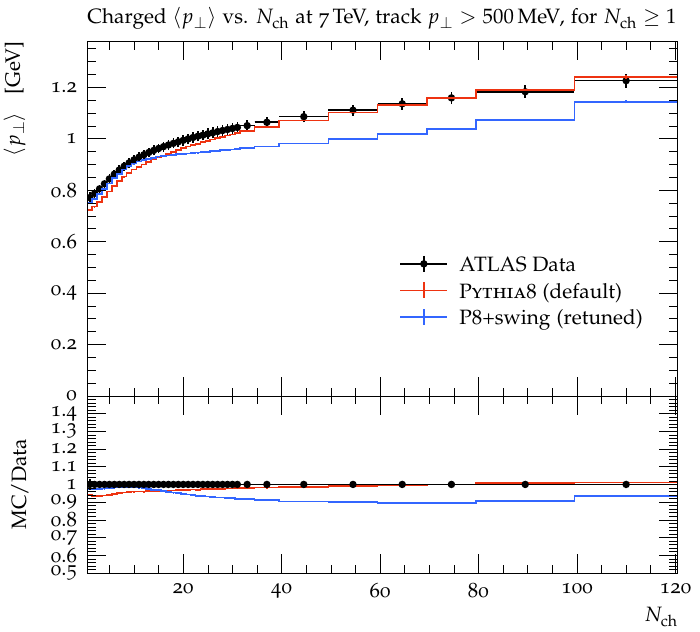}
  \end{minipage}%
  \caption[\pp\ event shape results]{%
    The effects of the \swing\ mechanism on \pp\ event shapes are here compared
    to ATLAS data, \cite{ATLAS:2010jvh}, and to the default \pythia setup.
    Minimum bias \pp\ events were generated at
    \mbox{$\sqrt{s} = 7$~TeV}.
    Top row: the pseudorapidity dependence of charged particles (left) and the
    distribution in charged event multiplicity (right).
    Bottom~row: the charged-particle $p_\perp$ distribution (left) and the mean
    transverse momentum as a function of the charged-particle
    multiplicity (right).
    All data comes from a sample obtained with a $p_\perp$ cut of 500~MeV.
    Note the cuts on $N_\text{ch}$ indicated for the top left plot.
  }
  \label{fig:ppres}
\end{figure*}

Although the value of $p_{\perp0,\text{ref}}$ needed to fit
multiplicities is uncomfortably high, the description of other
minimum-bias observables turns out to be rather good as illustrated
with some representative observables in \fig{ppres}. The only
troubling distribution is the average transverse momentum as a
function of the multiplicity of charged particles,
\mbox{$\langle p_\perp \rangle(N_\mathrm{ch})$}.
This observable is especially sensitive to colour reconnections and it is the very reason for introducing reconnections in \pythia when MPI were first
added~\cite{Sjostrand:1987su}.
The fact that we undershoot the data for high multiplicities is an indication that there is not enough colour reconnections in our \swing\ model.

We are aware of the fact that our \swing\ model is incomplete.
If we compare our approach to the QCD-based colour reconnection---which is also
based on dipole swings, albeit swings being applied only after all perturbative
emissions---we note that one important aspect is missing:
namely, reconnections that produce junction structures.
In principle, such reconnections can be implemented in our \swing\ model also
but the spacetime configuration of these junctions structures is very
complicated, and, as we will elaborate further on in \sec{disc}, we
will address these complications in a future publication.

\subsection{pPb results}
\label{sec:pPbres}
  With \pp\ multiplicity data reasonably well explained, we expect the new
\swing\ mechanism to integrate well with the generalised approach used
by the heavy-ion module in \pytppp in order to handle collision systems of
varying sizes.
Simulations of pPb collisions were undertaken and event properties were studied. 
The set of commands and parameter values used for the simulation setup is
provided in \app{pPb-setup}.
Compared to the setup applied to \pp\ collisions as of \app{pp-setup}, apart
from the obvious changes in beam setup commands, the configuration of pPb
simulation runs were very similar.

However, the heavy-ion machinery in the Angantyr model introduces some
parameters used to handle multiple NN subcollisions and their
merger into a single collision system.  The main issue when treating
collisions involving ions is the question
of how to handle cases where the proton beam interacts with more than
one nucleon inside the nucleus of the other beam.
Angantyr determines internally not only which nucleon(s) the proton interacts
with but also \emph{how} it interacts: elastically or inelastically.
For an inelastic subcollision, Angantyr further determines if it is
diffractive or non-diffractive.  Multiple elastic and diffractive
inelastic collisions are fairly easily dealt with but how do we model
cases where the proton interacts non-diffractively with several
nucleons in the nucleus?

The procedure as it is implemented in Angantyr is to take one such sequence
of non-diffractive proton--nucleon collisions and generate a single
non-diffractive NN subevent, while any subsequent non-diffractive
proton--nucleon collisions are generated \textit{as if} the nucleon is
only diffractively excited.  In \pythia's normal pp machinery, a
diffractive excitation is treated as a non-diffractive
\textit{Pomeron}--nucleon collision in which the Pomeron is emitted from the
proton according to a flux-function on the form
\mbox{$\mathrm{d}P(x_\pom) \propto \mathrm{d}x_\pom/x_\pom^{(1-\varepsilon)}$}
which in turn defines the mass distribution of the diffractively
excited nucleon as
\mbox{$\propto \mathrm{d}M_x^2/M_x^{2(1-\varepsilon)}$}.
However, in Angantyr, the generation of these secondary
non-diffractive subevents is done somewhat differently to a normal
diffractive event in \pythia so that, in the nucleon direction, the subevent
\emph{appears} more like a non-diffractive proton--nucleon event.  The
procedure to achieve this is described in detail in~\cite{Bierlich:2018xfw},
where it was argued that, contrary to standard diffractive excitation
modelling where $\varepsilon$ is small but positive, when used to model
secondary non-diffractive events, $\varepsilon$ should rather be
negative.

We will here use a recent modification to Angantyr in which secondary
non-diffractive subevents are generated as non-diffractive
proton-nucleon ones but with a special cut so that if the remnants of
the proton is removed, the remaining partonic state again has a mass
\mbox{$\propto dM_x^2/M_x^{2(1-\varepsilon)}$}~\cite{9131895}.  This
method simplifies the generation while yielding results very similar
to those of Angantyr without this modification.

The procedure for treating secondary non-diffractive subcollisions in Angantyr, with or without this modification, introduces an additional parameter for pPb (and PbPb) that is not relevant to \pp\ collisions.

\begin{figure}[t]
  \centering
  \includegraphics[width=0.48\textwidth]{%
    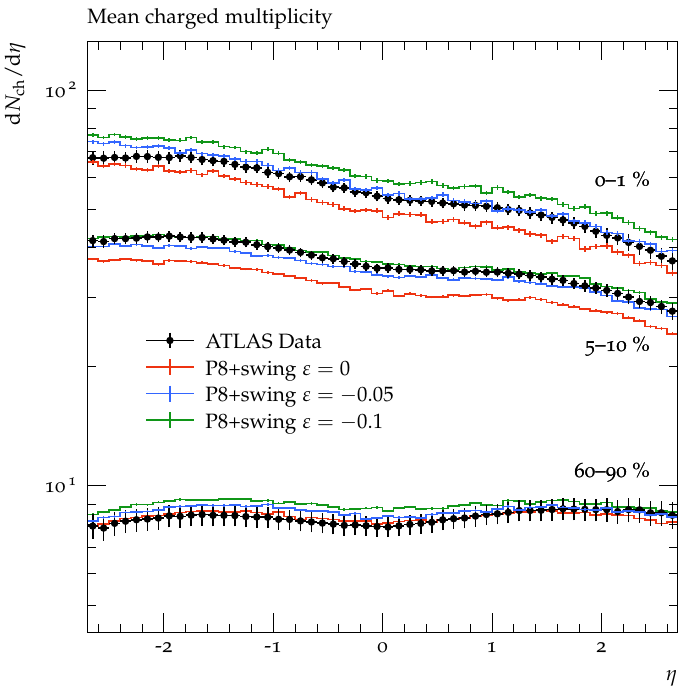}
  \hfill
  \includegraphics[width=0.48\textwidth]{%
    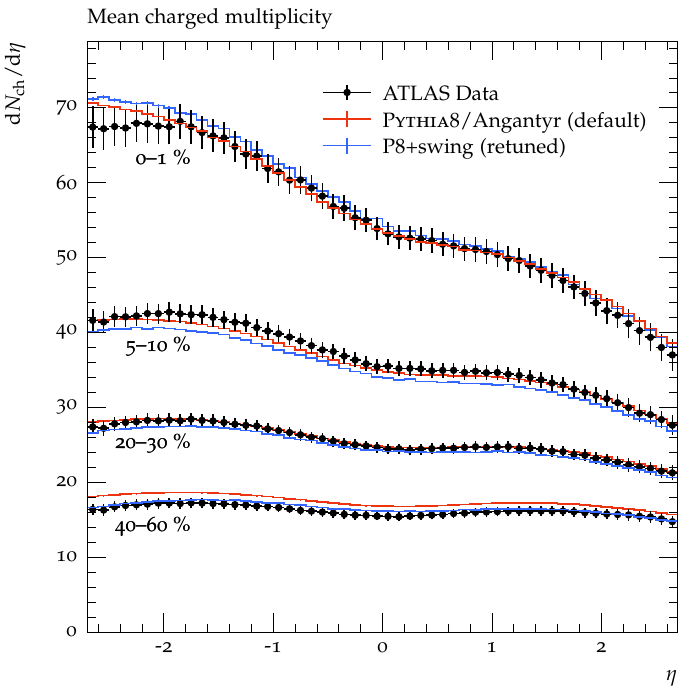}
  \caption[pPb tune]{%
    The plot displays the mean charged-particle multiplicity as a function of
    pseudorapidity in proton-lead collisions.
    Data for three different calibrated centrality bins, 0--1~\%, 5--10~\%, and
    60--90~\%, are presented and indicated with percentage ranges respectively
    (left).
    Simulation results of pPb events generated at
    \mbox{$\sqrt{s_\text{NN}} = 5.02$~TeV}
    for different values of the parameter $\varepsilon$ are compared to minimum
    bias ATLAS pPb data~\cite{ATLAS:2015hkr}.
    The comparison favours a choice of the parameter value of about -0.05.
    The plot on the right is the same data but on linear scale and a different
    choice of centrality bins compared to the retuned
    \mbox{($\varepsilon=-0.05$)}
    \swing\ model and to the default Angantyr setup.
  }
  \label{fig:pPbtune}
\end{figure}

For our new \swing\ model, we expect that the effect of allowing
reconnections between dipoles from all subevents in a pPb collision is to
significantly decrease the multiplicity, and this expectation holds in
particular for central events.
We therefore study the pseudorapidity distribution of charged
particles in different ``centrality'' bins.
As a reference, we use the ATLAS measurement in \cite{ATLAS:2015hkr}, where the centrality is
defined in terms of percentiles of the summed transverse energy in the
rapidity region
\mbox{$-4.9<\eta<-3.1$}
(close to the direction of the lead nuclei).
It was shown in \cite{Bierlich:2018xfw} that Angantyr
describes this data very well using
\mbox{$\varepsilon=0$}.

In \fig{pPbtune}, we show what happens when we include our new \swing\ model.
As expected, the multiplicity is decreased and for
\mbox{$\varepsilon=0$}
we undershoot data, especially for the most central
(0--1\% and 5--10\%) events.
However, we also see that, with a fairly
moderate adjustment of $\varepsilon$ from 0 to $-0.05$, we can
compensate for this and get as good a description of the data as is
obtained from the default Angantyr model.

\subsection{PbPb results}
\label{sec:PbPbres}
Given the very reasonable descriptions of multiplicity data for
\ee, \pp, as well as for pPb collisions presented in the
preceding results sections, our expectation at this stage was that we had
succeeded in constructing a colour reconnection model that would also
give a good description of experimental PbPb multiplicity data
without the need for any further tuning.
For the purpose of investigating
the validity of our expectation, PbPb collision events at
\mbox{$\sqrt{s_\text{NN}} = 2.76$~TeV}
were generated using our
modified and crudely tuned version of the Angantyr model (including the Gleipnir
code base) with the settings and tuned parameters from
the small-systems simulations discussed above.
No further tuning was attempted for PbPb collision simulations as there are no
additional parameters involved.
The set of commands and parameter values used for PbPb simulations, provided in
\app{PbPb-setup}, should therefore be compared to those for our pPb simulation
setup, see \app{pPb-setup}.

\begin{figure*}[t]
  \centering
  \includegraphics[width=0.48\textwidth]{%
    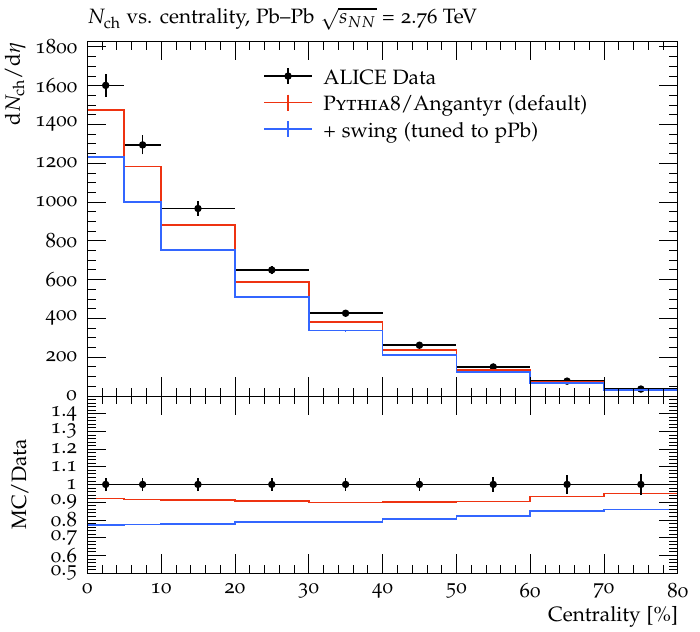}
  \hfill
  \includegraphics[width=0.48\textwidth]{%
    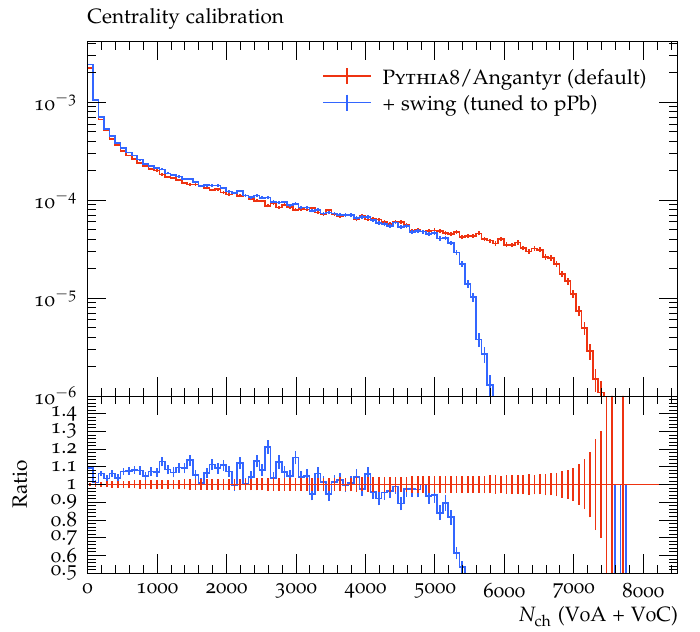}
  \caption[PbPb multiplicity results]{%
    Here, we provide a comparison to the ALICE results from \cite{ALICE:2010mlf}
    for the multiplicity at mid-rapidity in PbPb collisions at
    \mbox{$\sqrt{s_\text{NN}} = 2.76$~TeV}
    as a function of centrality, of the default Angantyr model,
    \cite{Bierlich:2018xfw}, and our new swing mechanism tuned to pPb data
    (left).
    Also shown is the centrality calibration for the generated samples, showing
    the summed charged multiplicity in the rapidity regions of the ALICE V0A and
    V0C detectors (right).
  }
  \label{fig:PbPbres}
\end{figure*}

In \fig{PbPbres} (left), we compare our new approach to the default 
Angantyr setup and to data from ALICE, \cite{ALICE:2010mlf}, of the
multiplicity at mid-rapidity as a function of centrality and it is evident that
we undershoot the data with the new \swing\ mechanism.
It may seem like we undershoot the multiplicity for all centralities but
it should be pointed out that the centrality binning is determined from
each set of event samples separately and this makes it difficult draw conclusions.

The centrality calculation as it is done for the ALICE experiment is a bit more
opaque compared to the one described above for the ATLAS experiment and cannot
directly be applied to our generated events.
As a proxy for the ALICE centrality estimation, we use the algorithm provided in
RIVET\footnote{%
  The analysis in RIVET is called
  \texttt{ALICE\_2015\_PBPBCentrality}.
}
where percentiles in the number of charged particles in the rapidity ranges of
their forward and backward detectors (a.k.a.\ V0A and V0C) defines the
centrality bins. The resulting distributions for the default Angantyr setup and
for our new \swing\ model is shown in \fig{PbPbres} (right).

Here, we see that the multiplicity distributions in the combined
forward-backward region are rather similar.
However, with our new swing model, the distribution dies out earlier for very
high multiplicities.
It is reasonable to assume that for very high multiplicities, the density of
dipoles is also very high.
Even though we expect the dipoles to be spread out
over a larger impact parameter area in central PbPb collisions
compared to the case of pPb, when triggering on high multiplicity, we
bias ourselves to very high densities where the probability of having
colour reconnections increases dramatically; the reconnections reduce the string
lengths, and, hence, also the multiplicity.
We believe that this is the reason for the multiplicity distribution dying out
much earlier in our new model.
This fact will be elaborated on further in our conclusions below.

\section{Discussion, conclusions, and outlook}
\label{sec:disc}
  %
In this work, we have presented a new colour reconnection algorithm
capable of reconnecting pairs of colour dipoles not only within
NN subcollisions but also between them, for the first
time accounting for the full spacetime picture of separation between
the partonic dipole ends.
This so-called \swing\ mechanism was designed to be interleaved with the
perturbative FSR shower and to be applied to any collision system.
In particular, it can be applied to collisions involving heavy ions and,
contrary to most other colour reconnection models, can act on the full partonic
state in order to provide a novel source of collective effects.

The algorithm itself is fairly complicated and it is trying to answer
questions such as:
where and when can a dipole be said to have formed;
when and where should it be allowed to \textit{swing} with another
dipole, irrespective of whether the other dipole belongs to a high
transverse momentum jet or to the soft underlying event;
if several swings are allowed, which one will happen \textit{first};
and would that happen before or after a gluon is emitted from one of the
dipoles?
There are no unique answers to some of these questions, and for others, the
answers that we did come up with are approximate at best.

The model as presented here is not yet complete and there is probably
still room for modifications in our answers to the questions above.
One thing that we clearly lack is the handling of so-called junction
structures.
So far, we have only dealt with the case of
dipole pairs where the quark in one dipole can form a singlet together with the
antiquark of the another, effectively cancelling the colour field between them
and the other dipole ends.
It is possible, also, to have dipoles where the two quarks may
form an antitriplet and giving rise to colour junction structures which
carry baryon numbers. The spacetime and energy-momentum structure
of the colour field connected to such junction is much more
complicated compared to the dual dipole structure and more work is needed before
such structures can be included in our model.

In this article, we have shown that the model as it stands works
approximately the way it is intended.
For \ee\ annihilation collisions, we see small effects, consistent with those
obtained by an earlier, purely momentum-based perturbative, swing model.
For \pp\ collisions, we find that the effects of the reconnections are a bit too small and that, even though multiplicities can be retuned, there is not enough
reconnections to properly describe the increase in average transverse
momentum with increasing event multiplicity found in experiments.
The inclusion of junction swings will surely increase the number of
reconnections but whether this sufficiently improves the transverse momentum
characteristics in our model remains to be seen.
In pPb collisions, we see a clear effect of reconnections happening between
different subcollisions, and this is one of the main motivations for
constructing the model.
These effects are even larger when considering PbPb collisions and, in fact,
they seem to be too large.
In the dense environment of central heavy-ion collisions, the probability
to reconnect and thereby minimising dipole invariant masses is so large that it
seems impossible to reach the high hadron multiplicity found in
experiments.
This issue would not be solved if more reconnections channels are introduced
with the junction swing.

Nevertheless, before giving up, we must acknowledge the fact that our \swing\
model is only one part of a set of models intended to give an
alternative to the conventional picture of a thermalised, hot, and
dense quark--gluon plasma being formed in heavy ion collisions. The
basic idea of these models is that also dense collision systems can be
described by the dynamics of colour fields between individual partons,
and our \swing\ only describes the first phase of this dynamics.
Later on in the collision process, as the distance between the colour charges
increases, the fields are confined in colour strings that can interact by repelling each other and this repelling force gives rise to flow.
Furthermore, when the strings eventually fragment into hadrons, the presence of overlapping strings will increase the string tension, allowing for more strange hadrons to be formed.
Finally, after hadrons have formed, there is the possibility
for them to interact with each other giving rise to further flow
effects~\cite{Sjostrand:2020gyg}.

Although those additional string interactions are not expected to give rise to
a considerable increase of multiplicities in dense systems, it has
already been shown that hadronic rescattering may do just that.
In \cite{Bierlich:2021poz}, it was found that hadronic rescattering
have a very large such effect, producing hundreds of extra particles
in the central rapidity region for very central events.

In this article, we have mainly been looking at multiplicities but
colour reconnections also influence many other observables and we
will study these in more detail in future publications.
What makes our model unique is that it influences the perturbative parton
evolution and we therefore expect interesting effects on, \eg, jet shapes.
Especially interesting is the effect on jets in heavy ion
collisions where we allow for swings between dipoles in a jet and
dipoles in the underlying event.

We note that there are some similarities between our swing model and
the so called anti-angular-ordering phenomena (see, \eg,
\cite{Mehtar-Tani:2011lic,Abreu:2024wka}).
A dipole between two partons inside a jet that has a small opening angle will
normally not radiate gluons at larger angles and this gives rise to the well
known principle of angular ordering in parton showers.
When a jet traverses a \textit{coloured medium}, however, one can expect that
the colours of the partons in the dipole may become decorrelated, allowing for
incoherent emissions of large-angle gluons.
The same effect can be achieved in our \swing\ model, where the dipoles in a
jet may swing with dipoles in the \textit{medium} (in our case, explicitly
modelled by soft partons in the underlying event) and thereby allowing for
large-angle FSR emissions.

In conclusion, we wish to emphasise that our microscopic approach to
particle collisions offers an interesting perspective on the treatment
of both small and large collision systems.  In recent years, the Lund group
has explored various ideas regarding the interactions between coloured
dipoles as well as between Lund strings, implementing them in multiple
explorative models.  These mechanisms are capable of producing several
phenomena relevant to the heavy-ion community and the study of
collectivity, including strangeness enhancement, flow, and, now with
this work, jet modification and medium response.  The extent to which
the Lund mechanisms, when considered collectively, can quantitatively
reproduce QGP-like signals remains to be studied but we recognise the
importance of discussing the role of microscopic models in the field
of heavy-ion physics and remain excited about furthering this
discussion with the community.

\section*{Acknowledgements}
\label{sec:acc}
  %
This work was partly funded by the Knut and Alice Wallenberg
foundation, contract number 2017.0036, and by the Swedish Research
Council, contract number 2020-04869.

\appendix

\section*{Appendices}

\section{A detailed overview of the \swing\ algorithm}
\label{sec:details}
  %
For the interested reader, this section provides a detailed overview
of the implementation of our \swing\ mechanism.  We begin by briefly
mentioning the relevant (\texttt{C++}) classes and objects that form
the foundation of the implementation.  Then, we walk through the
algorithm in detail, explaining its steps and logic.
Finally, the veto algorithm that, in the end, generates the proper time of a
swing for a dipole pair is described.

It should be noted that the simulation code written in preparation for
this paper and outlined below is not publicly accessible per se but
anyone who is interested in what we have done or how we implemented
our ideas is welcome to contact us.
We also point out here that, due to the experimental status of the code
discussed here, the code is expected to evolve further.
Nevertheless, we wish to provide this detailed
overview as to document the way we have addressed the challenges faced
during the implementation of our particular version of a dynamic colour
reconnection mechanism.

The code that implements our \swing\ mechanism has been added into an
existing code infrastructure on a developer branch of \pytppp, the
purpose of which is to manage several mechanisms that implement ideas
related to the emergence of collective behaviour within the Lund
string framework.
  
As part of any event setup during simulation, a new class has been
introduced in the code that handles not only the \swing\ mechanism but
also the mechanisms of shoving and rope hadronisation.
This new class \texttt{Gleipnir} is a pending replacement class for
the currently implemented \texttt{Ropewalk} class, the latter being
the publicly available implementation of the \emph{Rope Hadronization
framework}, \cite{Bierlich:2014xba}, in \pytppp.  The
\texttt{Gleipnir} class contains many helper classes among which the
most important to this work is the \texttt{GleipnirOverlap} class.  A
\texttt{GleipnirOverlap} object is to be set up for, a priori, every
combination of two \mbox{\texttt{GleipnirDipole}s} that may reconnect,
although we note that some dipole pairs need not to be considered due
to restrictions imposed by the \swing\ algorithm, \eg, mismatching
colour states.  The \texttt{GleipnirDipole} class represents a colour
dipole in the Gleipnir model.  At the beginning of the FSR shower, an
overlap object is initialised for all allowed pairwise combinations of
dipoles that has resulted from the ISR and the (semi)hard
scattering(s).  At present, this requires explicit separation of FSR
from ISR and MPI generation.

Each overlap object created is initialised so as the following
principles are adhered to.
Only dipole pairs with the same colour state are considered able to reconnect. 
A lower cut is imposed on the invariant mass of the two dipoles respectively in
order to avoid numerical issues when computing the parallel frame.

If these checks are passed, each overlap object computes the parallel
frame of its two dipoles to be used for reconnection calculations.
One more check is here performed to catch any case when the two
overlapping dipoles share a \texttt{GleipnirDipoleEnd}; a reconnection
of such a pair would result in a colour singlet (gluon) and this is
not permitted.
An object of the \texttt{GleipnirDipoleEnd} class represents to the Gleipnir
model a parton (or a junction) located at one end of a dipole.
If the parallel frame can be computed without
numerical issues, the overlap object stores the two angles that
characterise the frame (the relevant dipole opening angles are
visualised in the main text, see \fig{pf}).  Each overlap also stores
the transformation matrix that boosts a four-vector from the lab frame
to the parallel frame of the two dipoles.  It is noted whether the two
dipoles are \emph{parallel} or \emph{antiparallel}, \ie\ if the colour
charge with momentum $q_3$ and the anticolour charge with momentum
$q_4$ have the sign of their $x$- and $y$-components flipped w.r.t to
\eq{pfmoms}.

Following a successful computation of the parallel frame, an attempt
is made to generate a scale at which the reconnection of the two
dipoles will occur.  This attempt may or may not be successful as to
generate a reconnection that will take place within the evolution of
the shower.
The generating algorithm takes the current scale of the shower to be the
earliest possible scale at which a reconnection may happen.
However, since the shower is generated in order of
decreasing $p_\perp$ and our colour reconnections are to be ordered in
increasing proper time of the dipoles, the proper time $\tau$ passed to our
generating algorithm is
\mbox{$\tau = \hbar/p_\perp$} (together with an overall ``nudge'' factor in
order to control the translation between $\tau$ and $p_\perp$).
A simple check of the
argument is performed to ensure that the generation begins at
\mbox{$\tau < \tau_\text{S}$}.
For any larger $\tau$ passed to the
algorithm, the overlap is considered unable to reconnect.

In order for the algorithm to further determine if any given overlap
has a possibility to reconnect, several checks of overlap properties
are performed according to the principles outlined in the main text
before the actual attempt of generating a reconnection scale is made.
The algorithm makes use of the boost matrix stored in the overlap
object to boost production vertices and momenta of the four
dipole ends from their stored lab-frame values to the parallel frame
of the two dipoles of the overlap in question.
It is here worth mentioning that we have slightly modified the way vertices are
set for a non-Gaussian smearing mode, a mode that is first and foremost intended
for use in \ee\ collisions, in which the daughter parton(s) are assigned the
production point at the position their mother has at the time of
emission.\footnote{%
  The desired vertex assignment
  procedure can be set in our code through the parameter
  \texttt{PartonVertex:modeRadiation}.
}
Our modification entails
setting not only the three-space coordinates of the daughter parton(s)
but also the time of production.
From the boosted parton properties, the approximated dipole production vertex
$v_{ij}$ as defined in \eq{vclose} and the initial separation $\tau_{ij0}$ that
appears in \eq{taubar} are calculated, the latter being half the physical three
space separation at the point of closest approach.
A cut of a maximum of
$\tau_\text{S}$ is imposed on the $\tau_{ij0}$ for the secondary
dipoles so as to limit the closest separation to reasonable
values.\footnote{%
  The cut on $\tau_{ij0}$ for the secondary dipoles can
  be modified by a factor in our code through the parameter
  \texttt{Gleipnir:swingCut} which defaults to 1.0.
}
This cut
prohibits reconnections that would form one or two dipoles of which
the initial separation is so large that it has reached the
hadronisation scale.

The generation algorithm must identify any valid interval(s) in proper
time $\tau$ when the overlap may reconnect.  This is achieved using
calculations in the \emph{time} coordinate of the parallel frame and
the results are then cast into the proper time as given by the dipoles of
concern.

The minimum time $t_\text{min}$ after which a reconnection is
permitted is defined as the time at which either the two primary
dipoles or the two secondary dipoles have passed their respective
closest approach points as per \sec{restrictions}.  It is straight
forward to extract from the four approximated dipole production
vertices which dipole pair reaches this condition first by comparing
the times of the vertices as given in the parallel frame.  When
calculating the invariant proper time $\tau$ from the actual time $t$,
however, a complication arises due to the fact that all four dipoles
may well have originated at different spacetime points.  As was
outlined in \sec{ordering}, the furthest expanded dipole will be used
for calculating the ordering variable and this we take to mean that
when translating any time to a proper time as per \eq{taumax}, the
vertex time and velocity of the \emph{biggest} dipole will be used.
Biggest here refers to the dipole with the largest $\tau$ at the given
time of interest in the parallel frame and the observant reader should
note that which dipole this is, may change over the evolution and
expansion of the dipoles due to the fact that the primary and
secondary dipoles are moving with different velocities.  The minimal
time may, in the end, be used to find the minimal value
$\tau_\text{min}$ of the ordering variable $\tau$ for when the overlap
may reconnect.  The algorithm may adjust this lower bound upwards in
light of restrictions discussed further down.

No reconnection may be performed after the $\tau$ of the biggest
dipole has reached $\tau_\text{S}$ and this gives an intuitive initial
upper bound for the algorithm as
\mbox{$\tau_\text{max} = \tau_\text{S}$}.
The algorithm may adjust
this upper bound downwards in light of restrictions discussed further
down.

We argued in \sec{restrictions} that the dipoles must actually be in
colour contact for a reconnection to occur.  The dipoles are separated
in spacetime and are traversing the collision region.  The algorithm
must therefore identify the interval(s) in $\tau$ in which the primary
and secondary dipoles respectively have overlapping colour fields;
there are two such intervals in principle, one for each dipole pair.
This identification is done by simply evaluating the dynamic dipole separation
provided in \eq{separation} and finding the corresponding interval(s)
in $\tau$ when the biggest dipole has a \mbox{$\tau(t)$} that is
larger than the separation; this is interpreted as the dipoles being
spatially close and in colour contact.
The evaluation is done separately for the primary and secondary dipoles since
they are travelling and expanding in different directions.
The result of the colour contact calculation may restrict the proper time domain
in which a reconnection is permitted; there could be no valid interval, one
valid interval, or two valid intervals if the primary and secondary dipoles both
come into contact but their physical overlap is separated into two subintervals
with a \emph{hole} in between.

The above procedure identifies whether there exists a valid duration
in $\tau$ inside which a reconnection is permitted.
If any valid domain has been found, it is divided up into subintervals for
computational convenience over which the weight function of \eq{finalweight}
is strictly increasing or decreasing.
This is followed by assigning to each subinterval an \emph{overestimate} (OE) of 
the weight function to be used in a veto algorithm that attempts to generate a reconnection scale $\tau_\text{\swing}$ at which this overlap will compete with
the upcoming radiation from the shower.
The OE assigned to any subinterval is taken to be the largest value that the
weight function takes on that subinterval.
The value of the weight function includes the swing strength proportionality
constant\footnote{%
  This multiplicative constant can be modified in our code through the parameter
  \texttt{Gleipnir:SwingStrength} which defaults to 1.0.
}
mentioned in \sec{final-algorithm}.

If, after testing everything up to this point, the generating
algorithm has determined that there is at least one interval between
$\tau_\text{min}$ and $\tau_\text{max}$ during which a reconnection is
allowed, the incoming argument to the algorithm must be tested to
determine that it does not exceed $\tau_\text{max}$.
The algorithm then sets either the incoming argument or $\tau_\text{min}$ as the
starting point for generation depending on whichever one is largest.

The attempt at generating a reconnection scale can now be undertaken.
On the first subinterval in $\tau$, a logarithmic step is generated
using the OE of the subinterval and a
pseudorandom number, $R$, generated from a flat distribution;
they combine into a factor on the form
\mbox{$R^{-\frac{1}{\text{OE}}}$} which multiplies the current scale
to generate a new.  If the step overshoots the end point of the
subinterval, the algorithm resets to the starting point of the next
subinterval to which a different OE has been assigned and it
tries another logarithmic step forwards in $\tau$.
If there are no more subintervals to consider after an overshoot, the algorithm
terminates without having found a scale for a reconnection and considers the
overlap unable to reconnect.

In case the logarithmic step stays within the borders of the
current subinterval, the weight function is computed at the new value
of $\tau$ at which the algorithm ends up after the step.  The value of
the weight function is then compared to the OE value of the
subinterval, the latter which is multiplied by another uniformly
distributed pseudorandom number.  This comparison functions as an
accept/reject condition of the logarithmic step and the generated
scale is accepted if the value of the weight function at the generated
scale is larger that the pseudorandom fraction of the OE.

In cases when the generated scale is accepted, the scale is stored by
the overlap object and a control flag is set to indicate that this
particular overlap has a generated scale for a reconnection and that the
overlap is permitted to compete with the shower.

In cases when the generated scale is not accepted, the algorithm
simply takes another logarithmic step and tries again.

For every single overlap that we consider, the above procedure is
undertaken and the overlap object with the lowest value of its
reconnection scale (or, rather, the highest value of its inverse) is
compared to the $p_\perp$-scale of the next emission from the shower.
Any reconnection of dipoles or emission of a parton from the shower will modify,
or create, a subset of our overlap objects and, so, any modified overlap objects
will be subjected to a re-evaluation by our generation algorithm.
This procedure is repeated up until the hadronisation stage/the end point of the
\swing\ procedure.

\section{Simulation setup configurations}
\label{sec:tables}
  %
This appendix presents, in order of system size, the parameter values
and settings used for simulating collisions of \ee, \pp, pPb, as well
as of PbPb.  Note that the new \swing model is not yet available in
the official \pythia distribution. Any interested reader is, however,
certainly welcome to contact the authors to gain access to the code used to
produce the plots in this article.

\subsection{\texorpdfstring{\ee\ setup}{e+e- setup}}
\label{sec:ee-setup}
Note that in order to enable the \swing\ mechanism, for technical reasons, the
FSR shower and hadronisation must be turned off in the generation of
events and then forced manually in the main program code following the
event generation, according to the following sketch:
\scriptsize
\begin{verbatim}
  Pythia pythia;

  // Set parameters and switches, according to table below.
  pythia.init();

  for ( ... ) {  // The event loop.
    pythia.next();
    pythia.event = pythia.process;
    pythia.forceTimeShower(0, pythia.event.size() - 1.0, 92.2/2.0);
    pythia.forceHadronLevel();
  }
\end{verbatim}
\normalsize
This is currently the way of applying the \swing\
mechanism in lepton collisions, \ie\ when not utilising the built-in
heavy-ion machinery available to generate hadron/ion collision events.

The following table shows the settings we used to generate the event
sample with indications on how they differ from the default values.
Note in particular the switching off of the parton shower and hadronisation as
well as and the use of a non-standard option for the vertex calculation.
\begin{center}
  \begin{tabular}{@{}lrr@{}}  \toprule
    Parameters	& \pytppp	& w/ \swing\ \\  \midrule
    \texttt{PDF:lepton}	& off	& off \\
    \texttt{Beams:idA}	& 11	& 11 \\
    \texttt{Beams:idB}	& -11	& -11 \\
    \texttt{WeakSingleBoson:ffbar2gmZ}	& on	& on \\
    \texttt{23:onMode}	& off	& off \\
    \texttt{23:onIfAny}	& 1 2 3 4 5	& 1 2 3 4 5 \\  \midrule
    \texttt{PartonLevel:all}	& -	& off \\
    \texttt{HadronLevel:all}	& -	& off \\ 
    \texttt{PartonVertex:setVertex}	& -	& on \\
    \texttt{PartonVertex:modeRadiation}	& -	& 2 \\
    \texttt{Gleipnir:perturbativeSwing}	& -	& on \\
    \texttt{StringInteractions:model}	& -	& 3 \\
    \texttt{Gleipnir:testmode}	& -	& 40 \\  \midrule
    \texttt{TimeShower:alphaSvalue}	& -	& 0.14 \\  \bottomrule
  \end{tabular}
\end{center}

\subsection{\pp\ setup}
\label{sec:pp-setup}
Since our swing model is integrated with the Angantyr code for heavy-ion
collisions, we generate also pp collisions using the heavy-ion machinery
\mbox{(\texttt{HeavyIon:mode = 2})}.
The main difference to the standard pp machinery is that, rather than generating
the hardest partonic scattering before generating the impact parameter
(that will influence further scatterings), the heavy-ion machinery first
generates the impact parameter using its Glauber model before generating the
hardest scattering.

In the following table, we detail the settings used for the generated
samples of pp collisions.
Note in particular that parton showers in hadronic decays are
switched off; this is due to technical reasons.
The typical example of
decay-induced showers is the decay of $\Upsilon$.  Introducing proper
handling of such showers is expected to be straightforward and the
anticipated effects from such a fix are small.  Note also that the
vertex information in \pytppp must be turned on and the default
MPI-based colour reconnection mode must be explicitly turned off.
\begin{center}
\begin{tabular}{@{}lrrl@{}}  \toprule
  Parameters	& \pytppp	& w/ \swing\ \\  \midrule
  \texttt{Beams:eCM}	& 7000.	& 7000. \\
  \texttt{SoftQCD:all}	& on	& on \\
  \texttt{ParticleDecays:limitTau0}	& on	& on \\
  \texttt{ParticleDecays:tau0Max}	& 9.	& 9. \\  \midrule
  \texttt{HeavyIon:mode}	& -	& 2 \\
  \texttt{HeavyIon:SigFitNGen}	& -	& 4 \\
  \texttt{Angantyr:globalFSR}	& -	& 1 \\
  \texttt{ParticleDecays:FSRinDecays}	& -	& off \\
  \texttt{PartonVertex:setVertex}	& -	& on \\
  \texttt{ColourReconnection:reconnect}	& -	& off \\
  \texttt{Gleipnir:perturbativeSwing}	& -	& on \\
  \texttt{StringInteractions:model}	& -	& 3 \\
  \texttt{Gleipnir:testmode}	& -	& 40 \\  \midrule
  \texttt{TimeShower:alphaSvalue}	& -	& 0.14 \\
  \texttt{MultipartonInteractions:pT0Ref}	& -	& 2.60 \\  \bottomrule
\end{tabular}
\end{center}

\subsection{pPb setup}
\label{sec:pPb-setup}
The table below details the settings used for generating the pPb event samples.
Note in particular the last three settings that enables the new way of handling
secondary non-diffractive NN sub-events (see \sec{pPbres}) and setting the
corresponding $\varepsilon$ variable for the slope of the Pomeron flux factor,
together with enabling a slightly simplified model for positioning such
secondary sub-event in impact parameter space
The effect of the latter change is very small.
\begin{center}
  \begin{tabular}{@{}lrrl@{}}  \toprule
    Parameters	& \pytppp	& w/ \swing\ \\  \midrule
    \texttt{Beams:idA}	& 2212	& 2212 \\
    \texttt{Beams:idB}	& 1000822080	& 1000822080 \\
    \texttt{Beams:eA}	& 4000.	& 4000. \\
    \texttt{Beams:eB}	& 1570.	& 1570. \\
    \texttt{Beams:frameType}	& 2	& 2 \\
    \texttt{SoftQCD:all}	& on	& on \\
    \texttt{HeavyIon:SigFitNGen}	& 4	& 4 \\
    \texttt{ParticleDecays:limitTau0}	& on	& on \\
    \texttt{ParticleDecays:tau0Max}	& 9.	& 9. \\  \midrule
    \texttt{Angantyr:globalFSR}	& -	& 1 \\
    \texttt{ParticleDecays:FSRinDecays}	& -	& off \\
    \texttt{PartonVertex:setVertex}	& -	& on \\
    \texttt{ColourReconnection:reconnect}	& -	& off \\
    \texttt{Gleipnir:perturbativeSwing}	& -	& on \\
    \texttt{StringInteractions:model}	& -	& 3 \\
    \texttt{Gleipnir:testmode}	& -	& 40 \\  \midrule
    \texttt{TimeShower:alphaSvalue}	& -	& 0.14 \\
    \texttt{MultipartonInteractions:pT0Ref}	& -	& 2.60 \\
    \texttt{Angantyr:SASDmode}	& -	& -1 \\
    \texttt{Angantyr:epsilonSAND}	& -	& -0.05 \\
    \texttt{Angantyr:impactShift}	& -	& 1 \\  \bottomrule
  \end{tabular}
\end{center}

\subsection{PbPb setup}
\label{sec:PbPb-setup}
Finally, we provide here the table of settings used to produce the PbPb
event samples.
As noted in~\sec{PbPbres}, the only thing that really differs from the settings
in \app{pPb-setup} is the setting of the second beam but also some settings
specific to the ATLAS data not needed for out ALICE samples.
\begin{center}
  \begin{tabular}{@{}lrrl@{}}  \toprule
    Parameters	& \pytppp	& w/ \swing\ \\  \midrule
    \texttt{Beams:idA}	& 1000822080	& 1000822080 \\
    \texttt{Beams:idB}	& 1000822080	& 1000822080 \\
    \texttt{Beams:eCM}	& 2760.	& 2760. \\
    \texttt{SoftQCD:all}	& on	& on \\
    \texttt{HeavyIon:SigFitNGen}	& -	& 4 \\  \midrule
    \texttt{Angantyr:globalFSR}	& -	& 1 \\
    \texttt{ParticleDecays:FSRinDecays}	& -	& off \\
    \texttt{PartonVertex:setVertex}	& -	& on \\
    \texttt{ColourReconnection:reconnect}	& -	& off \\
    \texttt{Gleipnir:perturbativeSwing}	& -	& on \\
    \texttt{StringInteractions:model}	& -	& 3 \\
    \texttt{Gleipnir:testmode}	& -	& 40 \\  \midrule
    \texttt{TimeShower:alphaSvalue}	& -	& 0.14 \\
    \texttt{MultipartonInteractions:pT0Ref}	& -	& 2.60 \\
    \texttt{Angantyr:SASDmode}	& -	& -1 \\
    \texttt{Angantyr:epsilonSAND}	& -	& -0.05 \\
    \texttt{Angantyr:impactShift}	& -	& 1 \\  \bottomrule
  \end{tabular}
\end{center}

\bibliographystyle{utphys}
\bibliography{swing.bib}

\end{document}